\documentclass[12pt, twoside]{article}
\usepackage{a4wide,amssymb,cite}
\usepackage{epsfig}
\usepackage[usenames,dvipsnames]{color}
\usepackage{slashed}

\newcommand{\be}{\begin{equation}}
\newcommand{\ee}{\end{equation}}
\newcommand{\bea}{\begin{eqnarray}}
\newcommand{\eea}{\end{eqnarray}}

\begin{document}

\color{black}
\begin{flushright}
CERN-PH-TH/2012-239\\
KIAS-P12057
\end{flushright}

\vspace{1cm}
\begin{center}
{\huge\bf\color{black} Axion-mediated dark matter \\[3mm] and Higgs diphoton signal }\\
\bigskip\color{black}\vspace{1.5cm}{
{\large\bf Hyun Min Lee$^{a,b}$, Myeonghun Park$^{a}$ and Wan-Il Park$^{b}$}
\vspace{0.5cm}
} \\[7mm]

{\em $(a)$ {Theory Division, Physics Department, CERN,  CH--1211 Geneva 23,  Switzerland.}}\\
{\it $(b)$ School of Physics, KIAS, Seoul 130-722, Korea.  }\\
\end{center}
\bigskip
\centerline{\large\bf Abstract}
\begin{quote}\large
We consider axion-mediated dark matter models motivated by Fermi gamma ray line at $130\,{\rm GeV}$, where anomaly interactions of an axion-like scalar mediate a singlet Dirac fermion dark matter (DM) to electroweak gauge bosons.  In these models, extra vector-like leptons generate anomaly interactions for the axion and can also couple to the SM Higgs boson to modify the Higgs-to-diphoton rate. We can distinguish models by the branching fraction of the DM annihilation into a photon pair, favoring the model with a triplet fermion. From the condition that the lighter charged extra lepton must be heavier than dark matter for no tree-level DM annihilations, we also show that the ratio of Higgs-to-diphoton rate to the SM value is constrained by vacuum stability to $1.4(1.5)$ for the cutoff scale of $10(1)\,{\rm TeV}$.

\end{quote} 

\thispagestyle{empty}

\normalsize

\newpage

\setcounter{page}{1}

\section{Introduction}

Both ATLAS and CMS \cite{july4} have recently discovered a Higgs-like boson with mass $125-126\,{\rm GeV}$ with high significances of $5.9\sigma$ and $5.0\sigma$, respectively.  Although the overall signal strength is consistent with what we expected from the SM Higgs boson,  the Higgs-to-diphoton channel shows a larger signal strength than the Standard Model (SM) value.
Since the Higgs decay rate into a photon pair is induced by loops in the SM, there might be a room for new particles to play a role in modifying the Higgs couplings to a photon pair without having been ruled out by collider limits and electroweak precision test yet.

On the other hand, from the Fermi LAT data \cite{fermilat,fermilat2,fermilat3,fermilat4}, there is an interesting observation of the gamma-ray line at $130\,{\rm GeV}$ \cite{weniger}, which might be interpreted as the signal of dark matter annihilating into monochromatic photon(s) at the galactic center \cite{ibarra,weniger,raidal,su}. There are a lot of recent activities in building the dark matter models \cite{newmodels,lp2,tait} for explaining the Fermi gamma-ray line\footnote{See also other interpretations in Ref.~\cite{background}.} with satisfying the bounds on other annihilation channels \cite{pamela,fermilat3,continuum}. Seemingly unrelated two phenomena with di-photons from the Higgs boson decay and those from dark matter annihilation could find the same origin beyond the SM. This is the main topic of this paper, addressing the interplay between dark matter and Higgs signals in constraining the model parameters and the signal strength expected from the models.

We consider a microscopic theory of axion-mediated dark matter\cite{lp2}, which has been recently proposed by ourselves to accommodate the Fermi gamma-ray line with a Dirac fermion dark matter. 
To that purpose, we introduce extra vector-like leptons to generate the Peccei-Quinn(PQ) anomaly interactions of an axion-like pseudo-scalar to the SM electroweak gauge bosons. Then, the axion can mediate dark matter to annihilate into monochromatic photon(s) with the branching fraction depending on the representations of the extra leptons. Since we need a large coupling of the extra leptons to the Higgs boson to enhance the Higgs-to-diphoton rate \cite{wagner,AH,almeida,weiner}, 
the minimal content of extra leptons that we consider contains: (1) one vector-like pairs of doublets and a vector-like pair of singlets or (2) a vector-like pair of doublets and a triplet. 
We need to introduce two Higgs doublets to allow for the Higgs Yukawa couplings to the extra leptons. 
However, we focus on the decoupling limit of extra Higgs bosons where only the SM Higgs boson is kept in the low energy theory.

We show that Fermi gamma-ray line can be explained by the DM annihilation into a photon pair in both models. But, the two models are distinguishable by the different branching fractions of the DM annihilation cross section into a photon pair. When the lighter extra charged lepton state is smaller than around $200\,{\rm GeV}$,  the branching fraction of two-photon line increases significantly, as compared to the value determined by anomalies in the decoupling limit of extra lepton masses. 
In the singlet models, the branching fraction of two-photon line is large so extra annihilation channels with the scalar partner of the axion are needed.  On the other hand, in the triplet model, the branching fraction of two-photon line is consistent with the correct Fermi gamma-ray line and thermal relic density. 

We also show that the two models with extra leptons lead to the same Higgs-to-diphoton rate for the same charged fermion masses and couplings because the extra charged leptons are the same.  However, vacuum stability conditions become stronger for the triplet model in the region with the enhanced Higgs-to-photon rate. In order to make the loop-induced DM annihilation into monochromatic photon(s) sizable for the Fermi gamma-ray line, we need to suppress or forbid the tree-level DM annihilation into a pair of the lighter state of extra charged leptons.
Due to the nature of the axion coupling to the extra leptons, the phase-space suppression of the tree-level annihilation channel is not significant near where the lighter charged lepton is lighter than dark matter. 
Therefore, we need to forbid the tree-level channel kinematically by taking the lighter charged lepton mass to be larger than dark matter mass, $130\,{\rm GeV}$. In turn, combined with vacuum stability conditions, we obtain the bounds on the ratio of Higgs-to-diphoton rate to the SM value as $R_{\gamma\gamma} < 1.4(1.5)$ for the cutoff scale of $10(1)\,{\rm TeV}$.

The paper is organized as follows. We first present models with extra leptons and their interactions to the axion and the Higgs boson for the extended Higgs sector with two Higgs doublets and one complex singlet scalar. Then, we discuss the implications of extra leptons on the DM annihilation cross section into monochromatic photon(s)  and the Higgs-to-diphoton rate, impose the vacuum stability and perturbativity conditions, and electroweak precision constraints, and emphasize the interplay between dark matter and Higgs signals. 
Finally, the conclusions are drawn.

\section{Models with extra leptons}

We consider the extension of the SM with renormalizable interactions for axion-like scalar field $a$, an extra vector-like lepton $f$ and a singlet Dirac fermion dark matter $\chi$, as follows, 
\bea
{\cal L}&=&{\bar \chi}\gamma^\mu\partial_\mu \chi +{\bar f}\gamma^\mu D_\mu f+\frac{1}{2}(\partial_\mu a)^2 \nonumber \\
&&- m_\chi {\bar \chi} \chi- m_f {\bar f} f -\frac{1}{2}m^2_a a^2+\frac{1}{\sqrt{2}}i\lambda_\chi a {\bar \chi} \gamma^5 \chi + i\lambda_f a {\bar f} \gamma^5 f   \label{genlepton}
\eea
where $D_\mu$ is the covariant derivative with respect to the SM electroweak gauge group. 

When the extra leptons carry electromagnetic charges, they induce the effective interaction between the axion and two photons by triangle one-loops. 
Consequently, due to the direct axion coupling to the fermion DM, monochromatic photons can be generated from the DM annihilations mediated by the axion. 
For the large extra lepton mass $m_f=\lambda_f v_s$ with $v_s$ being a singlet vev,  PQ anomalies lead to the axion couplings to the electroweak gauge bosons,  
\be
{\cal L}_{a,{\rm eff}}=\sum_{i=1,2} \frac{c_i\alpha_i}{8\pi v_s}\, a F^i_{\mu\nu} {\tilde F}^{i\mu\nu} \label{anomalies}
\ee
where $c_1, c_2$ are constant parameters fixed as $c_1={\rm Tr}( q_{PQ}Y^2)$ and $c_2={\rm Tr}(q_{\rm PQ}l(r))$ with $q_{\rm PQ} (Y)$ being PQ charge (hypercharge) and $l(r)$ being the Dynkin index of representation $r$ of the extra lepton under the $SU(2)_L$.
For a pair of extra lepton doublets like Higgsinos in the MSSM,  we get $c_1=c_2=1$ for PQ charges normalized to $1$.

Furthemore, the extra vector-like lepton can have the effective interaction with the SM Higgs $H$ after heavy fermions are integrated out,
\be
{\cal L}_{H,{\rm eff}}= \frac{c_f}{M}H^\dagger H {\bar f} f. \label{effhiggs}
\ee
In the presence of multiple vector-like leptons $f_i(i=1,2,\cdots, N)$, the Higgs to di-photon decay rate can be enhanced or reduced, depending on the interference between the SM contribution and the new contributions with couplings $c_{f_i}$.

\subsection{Model I: vector-like doublet and singlet with PQ charges}

For concreteness, we consider PQ symmetry for a Dirac fermion dark matter that couples to the SM through the extra vector-like leptons.
We introduce a Dirac singlet dark matter fermion composed of $\chi, {\tilde \chi}$, and extra vector-like lepton doublets, $l_4,{\tilde l}_4$ with $l_4=(\nu_4,e_4)^T$ and ${\tilde l}_4=({\tilde e}_4,{\tilde \nu}_4)^T$, and extra vector-like lepton singlets, $e^c_4, {\tilde e}^c_4$. 
A complex scalar, $S$, mediates between dark matter and extra vector-like leptons while two Higgs doublets, $H_d, H_u$, are required to obtain the Higgs Yukawa couplings to extra leptons being consistent with PQ symmetry.

The interaction terms for dark matter and extra leptons are written  in terms of two-component Weyl spinors as follows,
\bea
-{\cal L}_{\rm Yukawa}
= \lambda_{\chi} S \chi {\tilde \chi}+ \lambda_l S l_4 {\tilde l}_4 + \lambda_e S e^c_4 {\tilde e}^c_4 + y_l H_d  l_4 e^c_4-  {\tilde y}_l H_u {\tilde l}_4 {\tilde e}^c_4+{\rm h.c.} \label{leptonY}
\eea
Here we note that the vector-like lepton doublet and singlet have the same PQ charges and the Higgs doublets have the same PQ charges as the one of the singlet $S$.
In the presence of PQ symmetry, the Yukawa couplings for quarks and charged leptons in the SM can be also written by assigning the PQ charges appropriately. 

For dimension-5 Majorana neutrino mass terms, $\frac{1}{M}(l_i H_u)(l_j H_u)$, where $M$ is the cutoff, we need to choose nonzero PQ charges for lepton doublets but zero PQ charges for lepton singlets. 
For Majorana neutrino case, the mixing term between the SM charged leptons and the extra charged leptons such as $H_d l_4 e^c_i$, $H_d l_i e^c_4$, are forbidden by PQ symmetry. Furthermore, the mixing dimension-5 terms between the extra heavy neutrino and the SM neutrinos are not allowed by PQ symmetry, so the lighter charged lepton would be stable in the minimal case. The PQ charges and $Z_2$ parities in Majorana neutrino case are given in Table~1.

\begin{table}[ht]
\centering
\begin{tabular}{|c||c|c|c|c|c|c|c|c|c|c|c|c|c|c|}
\hline 
& $q_i$ & $u^c_i$ & $d^c_i$ &  $l_i$ & $e^c_i$  & $l_4$ & $e^c_4$  & ${\tilde l}_4$ & ${\tilde e}^c_4$ & $H_u$ & $H_d$ & $S $ & $\chi$ & ${\tilde\chi}$ \\ [0.5ex]
\hline 
PQ & $1$ & $1$ & $1$  & $2$ & $0$ & $1$  & $1$  & $1$  & $1$  & $-2$  & $-2$ & $-2$ & $1$ & $1$\\ [0.5ex]
\hline 
$Z_2$ & $+$ & $+$ & $-$  & $+$ & $-$ & $-$  & $+$  & $+$  & $+$  & $+$  & $-$ & $-$ & $-$ & $+$\\ [0.5ex]
\hline
\end{tabular}
\caption{PQ charges and $Z_2$ parities for Majorana neutrino case}
\label{table:charges1}
\end{table}  

On the other hand,  if the PQ charges of lepton singlets are nonzero, only Dirac-type neutrino masses are allowed with right-handed neutrinos $N^c_i$. For Dirac neutrino case, the mixing term between the SM leptons and the extra vector-like leptons are allowed by the renormalizable Yukawa couplings so there should be an additional symmetry to protect large flavor violations in the lepton sector.
The PQ charges and $Z_2$ parities in Dirac neutrino case are given in Table~2.

\begin{table}[ht]
\centering
\begin{tabular}{|c||c|c|c|c|c|c|c|c|c|c|c|c|c|c|c|}
\hline 
& $q_i$ & $u^c_i$ & $d^c_i$ &  $l_i$ & $e^c_i$ & $N^c_i$ & $l_4$ & $e^c_4$  & ${\tilde l}_4$ & ${\tilde e}^c_4$ & $H_u$ & $H_d$ & $S $ & $\chi$ & ${\tilde\chi}$ \\ [0.5ex]
\hline 
PQ & $1$ & $1$ & $1$  & $1$ & $1$ & $1$ & $1$  & $1$  & $1$  & $1$  & $-2$  & $-2$ & $-2$ & $1$ & $1$\\ [0.5ex]
\hline
$Z_2$ & $+$ & $+$ & $-$  & $+$ & $-$ & $+$ & $-$  & $+$  & $+$  & $+$  & $+$  & $-$ & $-$ & $-$ & $+$\\ [0.5ex]
\hline
\end{tabular}
\caption{PQ charges and $Z_2$ parities for Dirac neutrino case}
\label{table:charges2}
\end{table}  

We note that in both Majorana and Dirac cases, Yukawa couplings are invariant under a shift in PQ charges, which reflects another accidental $U(1)_H$ global symmetry.

\subsubsection{Axion and Higgs couplings to extra leptons}

The PQ invariant potential for two Higgs doublets only is given by
\be
V_0(H_u,H_d)=\mu^2_1 |H_d|^2+\mu^2_2|H_u|^2 +\frac{1}{2}\lambda_1 |H_d|^4
+\frac{1}{2}\lambda_2|H_u|^4+\lambda_3 |H_u|^2 |H_d|^2+\lambda_4 |H_u H_d|^2.
\ee
The above Higgs potential $V_0$ possesses an additional global $U(1)_H$ symmetry.
Then, keeping the $U(1)_H$ in the singlet-Higgs couplings, we get the PQ and $U(1)_H$ invariant scalar potential for the singlet and the Higgs doublets,
\be
V(S,H_u,H_d)=\lambda_S |S|^4+ m^2_S |S|^2+\lambda_{H_u S} |S|^2|H_u|^2 +\lambda_{H_d S} |S|^2|H_d|^2 + V_0(H_u,H_d).
\ee
Thus, after PQ and $U(1)_H$ symmetries are broken spontaneously, there are two massless axions.
If the $U(1)$ symmetries are broken by the following soft mass terms,
\be
\Delta V= \frac{1}{2} m^{\prime 2}_S S^2 - \mu^2_3 H_u H_d +{\rm h.c.},
\ee 
the massless axions obtain masses but there is no mixing between the $S$ and Higgs axions. 
In both Majorana and Dirac neutrinos, the SM quarks are also charged under the PQ symmetry, so they could contribute to QCD anomalies through the Higgs Yukawa couplings. However, when two massless axions are lifted by the above $U(1)$ breaking soft masses,  there is no mixing between the Higgs pseudo-scalar and the singlet axion. So, the axion mediates dark matter interactions to electroweak gauge bosons through the couplings to extra vector-like leptons but it does not couple to the SM fermions or gluons directly. 

In the limit of decoupling the heavy additional Higgses, we can get the following effective potential containing the lighter CP-even Higgs (in unitary gauge) and the singlet only, 
\be
V_{\rm eff}=\lambda_S |S|^4 
+\mu^2_S |S|^2 +\Big(\frac{1}{2}m^{\prime 2} S^2+{\rm h.c.}\Big) + \frac{1}{2}\lambda_{hS} |S|^2 h^2+\frac{1}{4}\lambda_{h} h^4+\frac{1}{2}m^2_h h^2
\ee
where $\lambda_{hS},\lambda_{h}, m^2_h$ are the effective Higgs parameters written in terms of the input parameters in $V_{\rm tot}$.
For simplicity, henceforth we consider the effective theory with the heavy Higgses being integrated out. In this case, if there is no mixing between the lighter Higgs and the singlet, the lighter Higgs can be just like the SM Higgs.

After the singlet scalar and the Higgs doublet develop VEVs as $\langle S\rangle=v_s$, $H_d=\frac{1}{\sqrt{2}}(v_d,0)^T$ and $H_u=\frac{1}{\sqrt{2}}(0,v_u)^T$ with $v^2=v^2_u +v^2_d$, we get the mass terms,
\be
-{\cal L}_{\rm mass}=m_\chi \chi {\tilde\chi} + m_l(\nu_4 {\tilde \nu}_4+e_4 {\tilde e}_4) + m_e e^c_4 {\tilde e}^c_4+m_x e_4 e^c_4+ {\tilde m}_x {\tilde e}_4 {\tilde e}^c_4+{\rm h.c.}   
\ee
with $m_\chi=\frac{1}{\sqrt{2}}\lambda_\chi v_s$, $m_l=\lambda_l v_s$, $m_e=\lambda_e v_s$, $m_x=\frac{1}{\sqrt{2}} y_l v_d$ and ${\tilde m}_x= \frac{1}{\sqrt{2}} {\tilde y}_l v_u $. Then, the charged fermion mass matrix is
\be
({\tilde e}_4, e^c_4) \left(\begin{array}{ll} m_l & {\tilde m}_x \\ m_x &  m_e \end{array} \right)
\left( \begin{array}{l} e_4 \\ {\tilde e}^c_4  \end{array} \right).    \label{massmatrix}
\ee
Thus, the mass matrix squared is given by
\be
{\cal M}^\dagger_f {\cal M}_f=\left( \begin{array}{ll} m^2_l +m^2_x & m_l {\tilde m}_x+m_e m_x \\ m_l {\tilde m}_x+m_e m_x &  m^2_e+{\tilde m}^2_x  \end{array} \right).
\ee 
Then, after diagonalizing the squared matrix by a unitary transform with $V$,
\be
V^\dagger {\cal M}^\dagger {\cal M} \,V={\cal M}^2_D \label{diagonal}
\ee
with
\be
V=\left( \begin{array}{ll} \cos \theta_f & \sin\theta_f \\ -\sin\theta_f &  \cos\theta_f  \end{array} \right),
\ee
we get the mass eigenvalues and the mixing angle are
\bea
m^2_{f_{1,2}}=\frac{1}{2}\Big( m^2_l+m^2_e+m^2_x+{\tilde m}^2_x\mp \sqrt{(m^2_l+m^2_x-m^2_e-{\tilde m}^2_x)^2+4 (m_l {\tilde m}_x+m_e m_x)^2} \Big),  \label{eigenvalues} 
\eea
\bea
\sin (2\theta_f)=- \frac{2(m_l {\tilde m}_x+m_e m_x)}{m^2_{f_1}-m^2_{f_2}}. \label{mixing}
\eea
Accordingly, the Weyl fermion pair $(e_4,{\tilde e}^c_4)^T$ transforms as
\be
\left( \begin{array}{l} e_4 \\ {\tilde e}^c_4  \end{array} \right)=V\left( \begin{array}{l} f_1 \\ f_2  \end{array} \right) .  \label{rot1}
\ee
By performing a bi-unitary transform $U^\dagger {\cal M} \,V=M_D$ with
\be
U=\left( \begin{array}{ll} \frac{1}{m_{f_1}}(m_l\cos \theta_f-{\tilde m}_x \sin\theta_f) & \frac{1}{m_{f_2}}(m_l\sin\theta_f+{\tilde m}_x\cos\theta_f) \\ \frac{1}{m_{f_1}}(m_x\cos\theta_f-m_e \sin\theta_f) &  \frac{1}{m_{f_2}}(m_x\sin\theta_f+m_e\cos\theta_f)  \end{array} \right), 
\ee
we can diagonalize the charged fermion mass terms in eq.~(\ref{massmatrix}) as follows,
\be
m_{f_1} f_1 {\tilde f}_1+ m_{f_2 }f_2 {\tilde f_2}
\ee
Then, another pair of Weyl fermions, $({\tilde e}_4, e^c_4)^T$, transforms as
\be
\left( \begin{array}{l}  {\tilde e}_4\\ e^c_4  \end{array} \right)=U^*\left( \begin{array}{l} {\tilde f}_1 \\ {\tilde f}_2  \end{array} \right). \label{rot2}
\ee

Now we expand the singlets and Higgs doublets around the VEVs by $S=(v_s+s+ia)/\sqrt{2}$, $H_d=\frac{1}{\sqrt{2}}(v_d+h_d,0)^T$ and $H_u=\frac{1}{\sqrt{2}}(0,v_u+h_u)^T$. 
In the decoupling limit for the extra Higgs fields, the lightest CP-even Higgs becomes the SM Higgs while the lighter axion coming from $S$ remains in the effective theory.
Then, from the results in appendix A, we get the Yukawa couplings for the axion $a$ and the Higgs $h$,
\bea
-{\cal L}_{\rm int}&=& -i\frac{m_\chi}{v_s}\,a {\bar \chi} \gamma^5 \chi- i\frac{m_l}{v_s}\, a{\bar \nu} \gamma^5\nu\nonumber \\
&&-i a \,(\lambda_{1}  {\bar F}_1 \gamma^5 F_1+
\lambda_{2}  {\bar F}_2 \gamma^5 F_2) \nonumber \\
 &&+\frac{1}{2}ia\Big[(\lambda_{3}-\lambda_4) ({\bar F}_2 F_1-{\bar F}_1 F_2)-(\lambda_3+\lambda_4)({\bar F}_2 \gamma^5 F_1 +{\bar F}_1 \gamma^5 F_2) \Big]\nonumber \\
&&- y_1h   {\bar F}_1 F_1-y_2 h 
 {\bar F}_2 F_2 \nonumber \\
 &&-\frac{1}{2}h\Big[ (y_{3}+y_4)({\bar F}_2 F_1+ {\bar F}_1 F_2) +(-y_3+y_4)({\bar F}_2\gamma^5 F_1-{\bar F}_1 \gamma_5F_2)\Big]
 \label{yukawas}
\eea
where $\chi\equiv(\chi,{\tilde \chi}^{\dagger})$, $\nu\equiv(\nu_4,{\tilde \nu}^\dagger_4)$, $F_1\equiv(f_1,{\tilde f}^\dagger_1)^T$ and $F_2\equiv(f_2,{\tilde f}^\dagger_2)^T$ are the mass eigenstates in Dirac spinors.
For simplicity, we take $m_l=m_e$, $m_x={\tilde m}_x$ for which $\theta_f=\frac{\pi}{4}$ and $m_{f_{1,2}}=m_l\mp m_x$. 
Then, the Yukawa couplings are
\bea
\lambda_{1}&=&\frac{m_l}{v_s}=\lambda_{2}, \quad \lambda_{3}=\lambda_{4}=0,   \label{axionint1} \\
y_{1}&=&\frac{m_x}{v}=-y_{2},   \quad y_{3}=y_{4}=0.
\eea
In this case, eq.~(\ref{yukawas}) becomes
\bea
-{\cal L}_{\rm int}&=&- i\frac{m_\chi}{v_s}\,a\, {\bar \chi} \gamma^5 \chi- i\frac{m_l}{v_s}\, a\,{\bar \nu} \gamma^5\nu\nonumber \\
&&-i \lambda_{1} a \,(  {\bar F}_1 \gamma^5 F_1+
 {\bar F}_2 \gamma^5 F_2)
-y_1 h\,({\bar F}_1 F_1-{\bar F}_2 F_2).
\eea
On the other hand, using the results in appendix B,
we get the electroweak interactions of the extra leptons,
\bea
{\cal L}_{\rm gauge}&=&\frac{g}{2}W^\dagger_\mu \bigg[{\bar\nu} \gamma^\mu ( F_1 + F_2) +{\rm h.c.}\bigg] \nonumber \\
&&+\frac{e}{2\sin\theta_W\cos\theta_W}\,
Z_\mu\bigg[{\bar\nu}\gamma^\mu\nu+v_e({\bar F}_1\gamma^\mu F_1+{
\bar F}_2 \gamma^\mu F_2)+a_e({\bar F}_1 \gamma^\mu F_2+{\bar F}_2 \gamma^\mu F_1) \bigg] 
\nonumber \\
&&-e A_\mu ({\bar F}_1\gamma^\mu F_1+{\bar F}_2 \gamma^\mu F_2 )
\eea
where $v_{e}=-\frac{1}{2}+2\sin^2\theta_W$ and $a_e=-\frac{1}{2}$.
We note that there are also Yukawa couplings between the CP-even singlet scalar and dark matter/extra leptons. If they mix with the Higgs, they can affect the dark matter annihilation as well as the effective Higgs couplings to the SM. Here, we assume that the mixing between the CP-even singlet scalar and the Higgs is small enough not to affect the Higgs production cross section at the LHC.

\subsubsection{Dirac neutrino and $SU(3)_W$ unification}

For Dirac neutrino case, each pair of the SM lepton doublet and singlet also carries the same PQ charges as for the extra leptons, so the case is consistent with the $SU(3)_W$ unification \cite{su3w} in which the electroweak gauge group $SU(2)_L\times U(1)_Y$ is unified into a single one.
For instance, in 5D $SU(3)_W$ electroweak unification model on $S^1/(Z_2\times Z'_2)$ \cite{su3w2002}, $SU(3)_W$ is broken to $SU(2)_L\times U(1)_Y$ due to the orbifold boundary conditions, resulting in two fixed points on the orbifold:  $y=0$ where $SU(3)_W$ is unbroken and $y=\frac{\pi R}{2}$ with $R$ the radius of the extra dimension where $SU(3)_W$ is broken to $SU(2)_L\times U(1)_Y$. Thus, the SM leptons and extra leptons, that are a full $SU(3)_W$ triplet, are located at $y=0$ while the SM quarks, that are split multiplets of the $SU(3)_W$, are located at $y=\frac{\pi R}{2}$.  For the realistic Yukawa couplings for the charged leptons with $SU(3)_W$, the Higgs doublet must be extended to a bulk sextet, $H_{\bar 6}=(H_d, {\bar H}_T,{\bar H}_S)$, including a doublet $H_d$ with $Y=-1/2$, a triplet ${\bar H}_T$ with $Y=-1$ and a singlet ${\bar H}_S$ with $Y=-2$, instead of an $SU(3)_W$ Higgs triplet.  
There is one more bulk Higgs sextet, $H_6=(H_u,H_T, H_S)$, which contains another Higgs doublet, $H_u$. Then, the Higgs doublets survive as zero modes, giving rise to the Yukawa couplings for the SM quarks, leptons and extra leptons, while the extra Higgs states are projected out by orbifold boundary conditions.

In 5D $SU(3)_W$ unification on the orbifold, the KK towers of Higgs sextets living in bulk contribute to the running of the electroweak gauge couplings, Under the assumption that the unified electroweak gauge coupling becomes strong, we get the prediction on the Weinberg angle at $M_Z$ with KK corrections \cite{su3w2002},
\be
\sin^2\theta_W(M_Z)=0.25-\frac{3}{8\pi}\alpha_{\rm em} \Big[{\tilde B}\ln\frac{M_*}{M_c}+B\ln\frac{M_c}{M_Z}\Big]
\ee
where $\alpha(M_Z)=\frac{e^2}{4\pi}\simeq 1/128$,  $B=b_g-b_{g'}/3$ and ${\tilde B}={\tilde b}_g-{\tilde b}_{g'}/3$.
Here $b's$ denote the beta function coefficients for the SM zero modes 
and ${\tilde b}$'s denote those for the KK modes.
In the case with one pair of Higgs sextet and anti-sextet, the beta functions below the compactification scale are $b_g=3, b_{g'}=-\frac{21}{3}$ and $B= \frac{16}{3}$. Above the compactification scale, we get ${\tilde b}_g=-\frac{1}{3}$, ${\tilde b}_{g'}=-\frac{41}{6}$ and ${\tilde B}=\frac{35}{18}$. Then, from the observed value with $\sin^2\theta_W(M_Z)\simeq 0.231$, we can determine the unification and compactification scales under the strong coupling assumption with $M_*/M_c\sim 8\pi^3(100)$ as $M_*=140(77)\,{\rm TeV}$ and $1/R=1.7(2.4)\,{\rm TeV}$.

In the case of the $SU(3)_W$ unification, PQ anomalies respect the $SU(3)_W$ symmetry as follows,
\be
{\cal L}_{a,SU(3)_W}=\frac{c_0\alpha_0}{8\pi v_s}\, a F^W_{\mu\nu} {\tilde F}^{W\mu\nu} \label{su3anomaly}
\ee
with $\alpha_0\equiv \frac{g^2_0}{4\pi}$ with $g_0$ being the $SU(3)_W$ gauge coupling and $c_0=1$. Consequently, from the embedding of $U(1)_Y$ into $SU(3)_W$ with ${\tilde Y}=\frac{1}{\sqrt{3}}\,{\rm diag}(-\frac{1}{2},-\frac{1}{2},1)$ and $g_0  {\tilde Y}=g_1 Y$, we get the relation between the $SU(3)_W$ gauge coupling and the electroweak gauge couplings as $g_0=\sqrt{3} g_1=g_2$ at tree level. Thus, we obtain the tree-level value, $\sin^2 \theta_W=\frac{g^2_1}{g^2_1+g^2_2}=0.25$, at the $SU(3)_W$ unification scale.
Including the running of the gauge couplings below the unification scale, the electroweak anomaly couplings are given by the same formula (\ref{su3anomaly}) but in terms of electroweak gauge couplings,
\be
{\cal L}_{a,{\rm eff}}=\sum_{i=1,2} \frac{c_i\alpha_i}{8\pi v_s}\, a F^i_{\mu\nu} {\tilde F}^{i\mu\nu}
\ee
with $c_1=3 c_2$.  For an arbitrary vector-like pair of the $SU(3)_W$ representations, the ratio is fixed to $c_1/c_2=3$.  For a pair of $SU(3)_W$ triplets that we consider for minimality, we get $c_1=3$ and $c_2=1$.
The result is in contrast with the case where there are only a pair of extra lepton doublets like Higgsinos in MSSM for which $c_1=c_2=1$.
Furthermore, we also note that the $SU(3)_W$ requires a simpler mass matrix of extra charged leptons by the relation, $m_l=m_e$.

\subsection{Model II: vector-like doublet and PQ-neutral singlet}

We assume that a vector-like doublet lepton has nonzero PQ charges but a vector-like singlet lepton has zero PQ charges.   The corresponding interaction terms for dark matter and extra leptons are then the following,
\be
-{\cal L}=\lambda_\chi S \chi {\tilde\chi}+\lambda_l S l_4 {\tilde l}_4 +m_e e^c_4 {\tilde e}^c_4 +y_l H_d l_4 e^c_4 -{\tilde y}_l H_u {\tilde l}_4 {\tilde e}^c_4+{\rm h.c.}
\ee
where the vector-like singlet lepton has nonzero tree-level mass.

\begin{table}[ht]
\centering
\begin{tabular}{|c||c|c|c|c|c|c|c|c|c|c|c|c|c|c|}
\hline 
& $q_i$ & $u^c_i$ & $d^c_i$ &  $l_i$ & $e^c_i$  & $l_4$ & $e^c_4$  & ${\tilde l}_4$ & ${\tilde e}^c_4$ & $H_u$ & $H_d$ & $S $ & $\chi$ & ${\tilde\chi}$ \\ [0.5ex]
\hline 
PQ & $1$ & $0$ & $0$  & $1$ & $0$ & $1$  & $0$  & $1$  & $0$  & $-1$  & $-1$ & $-2$ & $1$ & $1$\\ [0.5ex]
\hline 
$Z_2$ & $+$ & $+$ & $-$  & $+$ & $-$ & $-$  & $+$  & $+$  & $+$  & $+$  & $-$ & $-$ & $-$ & $+$\\ [0.5ex]
\hline
\end{tabular}
\caption{PQ charges and $Z_2$ parities for Majorana neutrino case}
\label{table:charges3}
\end{table}

For Majorana neutrino case, the SM lepton doublets have the same PQ charges as the extra lepton doublets, so the mixing term between the SM leptons and the extra vector-like leptons are allowed by the renormalizable Yukawa couplings. The PQ charges and $Z_2$ parities are given in Table~3.

For Dirac neutrino case, we can forbid the Majorana neutrino mass term by choosing the PQ charges of the SM lepton doublets to be different from the extra lepton doublets. In this case, there is no mixing between the extra leptons and the SM leptons. The PQ charges and $Z_2$ parities including the right-handed neutrinos $N^c_i$ are given in Table~4.

\begin{table}[ht]
\centering
\begin{tabular}{|c||c|c|c|c|c|c|c|c|c|c|c|c|c|c|c|}
\hline 
& $q_i$ & $u^c_i$ & $d^c_i$ &  $l_i$ & $e^c_i$ & $N^c_i$ & $l_4$ & $e^c_4$  & ${\tilde l}_4$ & ${\tilde e}^c_4$ & $H_u$ & $H_d$ & $S $ & $\chi$ & ${\tilde\chi}$ \\ [0.5ex]
\hline 
PQ & $1$ & $0$ & $0$  & $0$ & $1$ & $1$ & $1$  & $0$  & $1$  & $0$  & $-1$  & $-1$ & $-2$ & $1$ & $1$\\ [0.5ex]
\hline
$Z_2$ & $+$ & $+$ & $-$  & $+$ & $-$ & $+$ & $-$  & $+$  & $+$  & $+$  & $+$  & $-$ & $-$ & $-$ & $+$\\ [0.5ex]
\hline
\end{tabular}
\caption{PQ charges and $Z_2$ parities for Dirac neutrino case}
\label{table:charges4}
\end{table}  

In the model with PQ-neutral extra singlets, the electroweak anomalies come only from the vector-like doublet lepton with anomaly coefficients, $c_1=c_2=1$.
Although the Higgs to diphoton decay rate is not changed as compared to the case with extra singlets with nonzero PQ charges, we get the branching fraction of DM annihilation into a photon pair to be about $14\%$ in the limit of large extra lepton masses as will be discussed in later section.

The Higgs interactions to extra leptons are the same as in Model I. But, the axion interactions are different from those in Model I because the singlet leptons do not couple to the axion. For $m_l=m_e$ and $m_x={\tilde m}_x$, the axion interactions in eq.~(\ref{yukawas}) become
\be
\lambda_1=\lambda_2=\lambda_3=\lambda_4=\frac{m_l}{2v_s}. \label{axionint2}
\ee

\subsection{Model III: vector-like doublet and triplet with PQ charges}

When the extra vector-like singlet fermion is replaced by a triplet fermion $T$ with zero hypercharge, the Higgs Yukawa couplings in eq.~(\ref{leptonY}) become
\be
-{\cal L}_{\rm Yukawa}=\cdots+y_l H_d l_4 T - {\tilde y}_l H_u {\tilde l}_4 T+{\rm h.c.} 
\ee
Then, as in the model with vector-like doublet and singlet leptons, there are two charged Dirac fermions with masses $m_{f_1}, m_{f_2}$. The effective Yukawa couplings for the axion and the lighter CP-even Higgs are obtained similarly as in eq.~(\ref{yukawas}).  We can assign the same PQ charge and $Z_2$ parity for the triplet fermion as for the vector-like singlet fermion.
Consequently, as far as the Higgs to diphoton rate is concerned, the triplet model is similar to Model I. 

The triplet model has crucial differences from Model I. First,  vacuum stability bounds are stronger in the triplet model due to a more negative contribution of the triplet fermion in the RG equation of the Higgs quartic coupling \cite{AH}.  Second, in the triplet model, the coefficients of the axion interactions to electroweak gauge bosons are given by
$c_1=1$ and $c_2=3$, in the limit of large extra lepton masses, in contrast to $c_1=3$ and $c_2=1$ in Model I.  Therefore, the intensity of the two-photon line is comparable to the one-photon line. 
In this case, the Fermi gamma-ray line 
can be still explained by the two-photon line, when dark matter mass is $130\,{\rm GeV}$ \cite{tait}.

\section{Interplay between dark matter and Higgs signal}

In this section, we consider various constraints on the model with vector-like leptons, from the DM annihilation cross section into monochromatic photons, the Higgs to diphoton decay rate and electroweak precision data as well as cosmology and collider bounds.

\subsection{DM annihilation into monochromatic photon(s)}

From the interaction between the axion and the vector-like charged extra lepton in eq.~(\ref{genlepton}), we get the analytic expression for the amplitude for the axion decay into a pair of photons, ${\cal M}=\epsilon^{*\mu}(k_1)\epsilon^{*\nu}(k_2){\cal M}_{\mu\nu}$, with
\be
{\cal M}_{\mu\nu}=i\,\frac{\lambda_f\alpha_{\rm em}}{\pi }\frac{ A_1(\tau_f)}{m_f}\epsilon_{\nu\rho\mu\sigma} k_2^\rho k_1^{\sigma}
\ee
where $\tau_f\equiv 4m^2_{f} / m^2_a$ and
\bea
A_1(\tau)&\equiv& \tau \arcsin^2( 1/\sqrt{\tau}).
\eea
In the limit of $\tau\gg 1$, $A_1(\tau)\approx1$, we obtain the following,
\be
{\cal M}_{\mu\nu}\approx i \frac{\lambda_f\alpha_{\rm em}}{\pi}\frac{ 1}{m_{f}}\epsilon_{\nu\rho\mu\sigma} k^\rho_2 k_1^{\sigma}.
\ee
Then, we can approximate the axion coupling to a pair of photons as 
\be
{\cal L}_{a\gamma\gamma}= c_{a\gamma\gamma} a F_{\mu\nu} {\tilde F}^{\mu\nu}
\ee
with
\be
c_{a\gamma\gamma}=\frac{\lambda_f\alpha_{\rm em}}{4\pi }\frac{ 1}{m_{f}}
\ee

\begin{table}[h!]\small
\begin{center}
\begin{tabular}{c||c|c|c}
 \hline
 & Model I & Model II & Model III
\\\hline\hline
$(c_1,c_2)$ & $(3,1)$ & $(1,1)$ & $(1,3)$
\\
${\rm Br}({\bar\chi}\chi \rightarrow \gamma\gamma)$ & $40\%$  &  $14\%$   & $6.5\%$
\\
${\rm Br}({\bar\chi}\chi \rightarrow WW)$ & $44\%$  &  $62\%$   & $65\%$
\\
${\rm Br}({\bar\chi}\chi \rightarrow ZZ)$ & $16\%$  &  $16\%$   & $15\%$
\\
$(r, R)$ & $(1.15\times 10^{-3},0.56)$  &  $(0.27,1.77)$   & $(1.01, 2.51)$
\\ 
\hline
\end{tabular}
\caption{Branching fractions for DM annihilation cross sections in the decoupling limit of extra leptons. Here, $r\equiv\langle\sigma v\rangle_{Z\gamma}/(2\langle\sigma v\rangle_{\gamma\gamma})$ and $R\equiv \langle\sigma v\rangle_{WW}/(2\langle \sigma v\rangle_{\gamma\gamma}+\langle \sigma v\rangle_{Z\gamma})$. }
\label{table:Br}
\end{center}
\end{table}

For the axion effective interactions to electroweak gauge bosons, we can take the similar limits of large fermion masses running in loops or equivalently compute the anomaly coefficients. Then, using the axion effective interactions, from Ref.~\cite{lp2}, we obtain the DM annihilation cross sections into $\gamma\gamma$, $Z\gamma$, $WW$ and $ZZ$, respectively, in terms of the anomaly coefficients, $c_1,c_2$, in eq.~(\ref{anomalies}),  
\bea
\langle\sigma v\rangle_{\gamma\gamma}&=&\frac{|\lambda_\chi|^2}{2048\pi^3 v_s^2}\,(c_1\alpha_1\cos^2\theta_W+c_2\alpha_2\sin^2\theta_W)^2\,{\cal S}, \\
\langle\sigma v\rangle_{Z\gamma}&=&\frac{|\lambda_\chi|^2}{4096\pi^3 v_s^2}\,(c_2\alpha_2-c_1\alpha_1)^2\sin^2(2\theta_W)\,{\cal S}\,\Big(1-\frac{M^2_Z}{4m^2_\chi}\Big)^3, \\
\langle\sigma v\rangle_{WW}&=& \frac{|\lambda_\chi|^2c_2^2 \alpha^2_2}{1024\pi^3 v_s^2}\,{\cal S}\,\Big(1-\frac{M^2_W}{m^2_\chi}\Big)^{3/2}, \\
\langle\sigma v\rangle_{ZZ}&=& \frac{|\lambda_\chi|^2}{2048\pi^3 v_s^2}\,
\,(c_2\alpha_2\cos^2\theta_W+c_1\alpha_1\sin^2\theta_W)^2\,{\cal S}\,\Big(1-\frac{M^2_W}{m^2_\chi}\Big)^{3/2}
\eea
where
\be
{\cal S}\equiv \frac{s^2}{(s-m^2_a)^2+\Gamma^2_a m^2_a}, \quad\quad s\simeq 4m^2_\chi.
\ee
Then, the ratio of the intensities of two-photon to one-photon lines is
\be
r\equiv \frac{\langle\sigma v\rangle_{Z\gamma}}{2\langle\sigma v\rangle_{\gamma\gamma}}=\frac{(c_2\alpha_2-c_1\alpha_1)^2\sin^2(2\theta_W)}{4(c_1\alpha_1\cos^2\theta_W+c_2\alpha_2\sin^2\theta_W)^2}\,\Big(1-\frac{M^2_Z}{4m^2_\chi}\Big)^3.
\ee
Therefore, we get the following results with $m_\chi=130\,{\rm GeV}$: in Model I with $c_1=3, c_2=1$, $r=1.15\times 10^{-3}$; in Model II with $c_1=c_2=1$, $r=0.27$; in Model III with  $c_1=1, c_2=3$, $r=1.01$.
In Model III, even if the intensity of two-photon line is comparable to the one of one-photon line, the best fit with two lines occurs for $m_\chi=130\,{\rm GeV}$ as in the two former models \cite{tait}. In Table~\ref{table:Br}, we summarize the branching fractions for DM annihilation cross sections in the decoupling limit of extra leptons.
The fraction of the continuum photons, $R\equiv \langle\sigma v\rangle_{WW}/(2\langle \sigma v\rangle_{\gamma\gamma}+\langle \sigma v\rangle_{Z\gamma})$, is also shown in the same table and we find that the continuum photons produced from the $WW$-channel would be consistent with the line spectrum in all the three models \cite{continuum}.

Now we consider the effect of the finite extra lepton masses in the DM annihilation cross sections.
The amplitude for the DM annihilation into a pair of photons is
\be
{\cal M}_{\chi {\bar \chi}\rightarrow\gamma\gamma}=({\cal M}_{\chi{\bar\chi}\rightarrow a})\,\left(\frac{i}{s-m^2_a-i\Gamma_a m_a}\right)\,({\cal M}_{a\rightarrow \gamma\gamma}) \label{matrix}
\ee
where $s$ is the center of mass energy squared and $\Gamma_a$ is the total decay width of the axion-like scalar. Then, the annihilation cross section is given by
\be
\langle\sigma v\rangle_{\gamma\gamma}=\frac{1}{16\pi s}|{\cal M}_{\chi{\bar\chi}\rightarrow \gamma\gamma}|^2. \label{xsection0}
\ee
Here, the squared amplitude for $a \rightarrow \gamma\gamma$ is obtained from the result of the decay width of the axion to two photons at one loop,
\bea
\Gamma_{a \rightarrow\gamma\gamma}&=&\frac{1}{32\pi m_a} |{\cal M}_{a \rightarrow \gamma\gamma}|^2 \nonumber \\
&=&\frac{|\lambda_f|^2\alpha^2_{\rm em}}{64\pi^3} \,m^3_a \left|\frac{A_1(\tau_f)}{m_{f}}\right|^2. \label{axiondecay}
 \eea
Since the axion-like scalar can be off-shell in the s-channel, we need to replace $m^2_a$ in eq.~(\ref{axiondecay}) with $s$ for the correct $|{\cal M}_{a\rightarrow \gamma\gamma}|^2$ in eq.~(\ref{xsection0}). Thus we get
\be
 |{\cal M}_{a \rightarrow \gamma\gamma}|^2=\frac{|\lambda_f|^2\alpha^2_{\rm em}} {2\pi^2} \,s^2 \left|\frac{A_1(\tau_f)}{m_{f}}\right|^2
\ee
where $\tau_f$ is now replaced by $4m^2_f/s$.
Therefore, plugging the above into eq.~(\ref{xsection0}) with eq.~(\ref{matrix}), we obtain 
\be
\langle \sigma v\rangle_{\gamma\gamma} =\frac{|\lambda_\chi\lambda_f|^2\alpha^2_{\rm em}}{512\pi^3}\frac{s^2}{(s-m^2_a)^2+\Gamma^2_a m^2_a}
\left|\frac{A_1(\tau_f)}{m_{f}}\right|^2.
\ee

\begin{figure}[t]
\centering
\includegraphics[width=7.5cm]{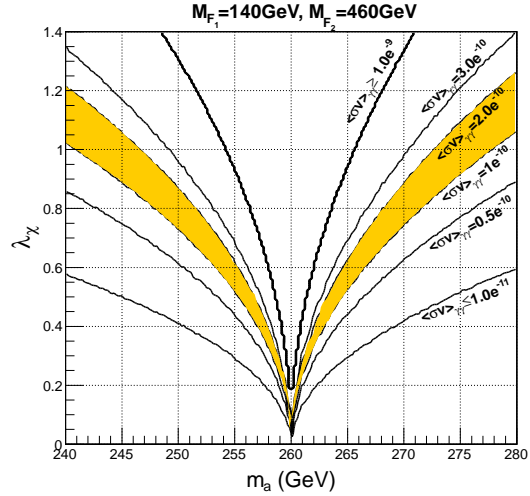} 
\caption{Parameter space of $m_a$ and $\lambda_\chi$ where the annihilation cross section into a photon pair in Model I. Yellow region is between the central values of the cross section for Einasto and NFW dark matter profiles, $10^{-10}\,{\rm GeV}^{-2}{\rm c}$ and $2\times 10^{-10}\,{\rm GeV}^{-2}{\rm c}$, respectively. We took ${\tilde y}_l=y_l/\tan\beta$ for $m_x={\tilde m}_x$ and $m_h=125\,{\rm GeV}$ for Higgs mass. We chose $m_{f_1}=140\,{\rm GeV}$ and $m_{f_2}=460\,{\rm GeV}$, 
which leads to $R_{\gamma\gamma}\simeq 1.4$ being consistent with vacuum stability at $1\,{\rm TeV}$. }
\label{fig:amass}
\end{figure}

In Model I, there are a pair of mass eigenstates of extra charged leptons with axion couplings, $\lambda_1, \lambda_2$, which are given in (\ref{axionint1}).  So, we get the DM annihilation cross section to a photon pair as
\be
\langle \sigma v\rangle_{\gamma\gamma} =\frac{|\lambda_\chi|^2\alpha^2_{\rm em}}{512\pi^3}\frac{s^2}{(s-m^2_a)^2+\Gamma^2_a m^2_a}
\left|\frac{\lambda_1 A_1(\tau_1)}{m_{f_1}}+\frac{\lambda_1 A_1(\tau_2)}{m_{f_2}}\right|^2
\ee
with $\tau_i\equiv 4m^2_{f_i}/s$.
On the other hand, from the results of appendix C, the DM annihilation cross section to $Z$-boson and one photon is
\bea
\langle\sigma v\rangle_{Z\gamma}&=&\frac{|\lambda_\chi|^2\alpha_{\rm em}\alpha}{1024\pi^3}\,\frac{(-0.5+2\sin^2\theta_W)^2}{\cos^2\theta_W}\frac{s^2}{(s-m^2_a)^2+\Gamma^2_a m^2_a}
\Big(1-\frac{m^2_Z}{4m^2_\chi}\Big)^3 \times \nonumber \\
&&\quad\times \bigg|\frac{\lambda_1 A_4(\tau_1,\rho^Z_1)}{ m_{f_1}} +\frac{\lambda_2 A_4(\tau_2,\rho^Z_2)}{ m_{f_2}}\bigg|^2
\eea
where $\rho^Z_i\equiv 4m^2_{f_i}/m^2_Z$ and 
\be
A_4(\tau,\rho^V) 
\equiv \frac{\tau \rho^V}{\tau -\rho^V}\left\{\arcsin^2\left(1/\sqrt{\rho^V}\right)-\arcsin^2\left(1/\sqrt{\tau}\right)\right\}.
\ee
Therefore, we get the ratio of two-photon to one-photon cross sections for Model I as
\be
\frac{\langle\sigma v\rangle_{Z\gamma}}{\langle \sigma v\rangle_{\gamma\gamma}}=\frac{\alpha(-0.5+2\sin^2\theta_W)^2}{2\alpha_{\rm em}\cos^2\theta_W}\,\Big(1-\frac{m^2_Z}{4m^2_\chi}\Big)^3 \,
\frac{\bigg|\lambda_1 A_4(\tau_1,\rho^Z_1)/ m_{f_1} +\lambda_2 A_4(\tau_2,\rho^Z_2)/ m_{f_2}\bigg|^2}{ \bigg|\lambda_1 A_1(\tau_1)/m_{f_1}+\lambda_2 A_1(\tau_2)/m_{f_2}\bigg|^2}.
\ee
So, for $\sin^2\theta_W(M_Z)\simeq 0.231$, the overall factor of the ratio is very small so the one-photon channel is negligible as the factor containing the loop functions is of order one. Then, it is possible to obtain the gamma ray line at $130\,{\rm GeV}$ from the two-photon peak, when dark matter mass is $130\,{\rm GeV}$. We note that the DM annihilation cross section into a pair of photons is required to be $\langle\sigma v\rangle_{\gamma\gamma}=(1.27\pm 0.32^{+0.18}_{-0.28})\times 10^{-27}{\rm cm}^3 {\rm s}^{-1}$ for the Einasto profile and
$\langle\sigma v\rangle_{\gamma\gamma}=(2.27\pm 0.57^{+0.32}_{-0.51})\times 10^{-27}{\rm cm}^3 {\rm s}^{-1}$ for the NFW profile, that is, ${\rm Br}({\bar\chi}\chi\rightarrow \gamma\gamma)\simeq 4-8\%$ for thermal dark matter \cite{weniger}. Therefore, since the DM annihilation into a photon pair is loop-suppressed, we need to rely on a resonance effect near $m_a\sim 2 m_\chi$ to get the correct annihilation cross section \cite{lp2}. In Fig.~\ref{fig:amass}, we depict the parameter space of the axion mass $m_a$ versus dark matter coupling $\lambda_{\chi}$, satisfying the required annihilation cross section into a photon pair.  We can see that the required tuning in the axion mass can be smaller for a larger dark matter coupling. For instance, for $\lambda_\chi\lesssim 0.8$, the axion mass can be deviated by $|m_a-2m_\chi|\lesssim10\,{\rm GeV}$ from the resonance mass.

\begin{figure}[t]
\centering
\includegraphics[width=16cm]{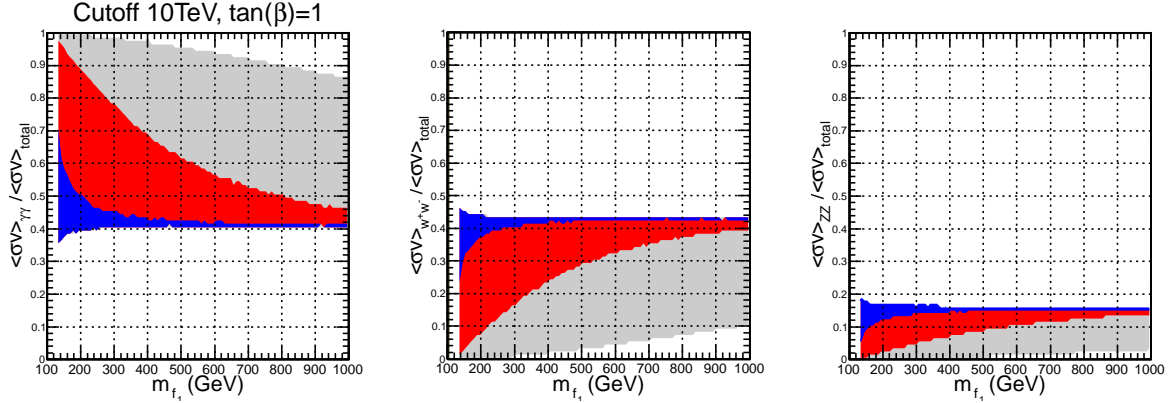} 
\caption{Branching fractions of the DM annihilation cross sections into a photon pair, $WW$ 
and $ZZ$, as a function of the lighter charged lepton mass in Model I. Gray region is for $m_{f_2}<10\,{\rm TeV}$ and red region is for $m_{f_2}<2\,{\rm TeV}$. We varied over $0<\lambda_\chi<1$ and took ${\tilde y}_l=y_l/\tan\beta$ for $m_x={\tilde m}_x$ and $m_h=125\,{\rm GeV}$ for Higgs mass. In red region, the annihilation cross section, $\langle \sigma v\rangle_{\gamma\gamma}=(1.27-2.27)\times 10^{-27}{\rm cm}^3 {\rm s}^{-1} $, is obtained. Blue region is consistent with perturbativity, vacuum stability and EWPD within $2\sigma$, which will be discussed in later sections.  }
\label{fig:rgg0}
\end{figure}

Similarly, as given in appendix C, we also obtain the matrix elements for the other channels such as the intermediate axion going to $ZZ$ and $WW$ for Model I. 
The $WW$-channel is induced by loops containing the axion couplings to the heavier charged lepton and the extra neutrino. On the other hand, charged fermion leptons and extra neutrino couple to the axion with similar strength, in particular, with the same strength for the maximal mixing of charged leptons. 
Thus, for a small mass of the lighter charged fermion, the branching fraction of the $WW$-channel is much smaller than the one of the channel with a photon pair, even if they are similar in the limit of large charged lepton masses. As shown in Fig.~\ref{fig:rgg0}, constraints coming from perturbativity, vacuum stability at $10\,{\rm TeV}$ and EWPD within $2\sigma$ are stronger than the one from Fermi gamma-ray line. In the parameter space consistent with all the constraints, the branching fraction of two-photon line depends on the lighter charged fermion mass: it becomes larger than $50
\%$ for $m_{f_1}<200\,{\rm GeV}$  while it approaches $40\%$ in the limit of $m_{f_1}\gg m_{\chi}$. Moreover, the $WW$ channel is next-to-dominant and it becomes smaller than $40\%$ for $m_{f_1}<280\,{\rm GeV}$ and it approaches $44\%$ in the large fermion mass limit.
Now we recall the $\gamma\gamma$-channel should occupy about $4-8\%$  branching fraction of thermal cross section for explaining the $130\,{\rm GeV}$ Fermi gamma-ray line, depending on the dark matter profile.
Thus, in order to get the correct branching fraction of the $\gamma\gamma$-channel in Model I,  there must be extra channels for DM annihilation at freezeout with at least $\langle\sigma v\rangle_X=(6.5-14)\langle\sigma v\rangle_a$ where $\langle\sigma v\rangle_a$ is the annihilation cross section from axion mediation.
The scalar partner of the singlet axion can give rise to extra DM annihilation channels into a pair of the SM particles through the Higgs-singlet coupling \cite{ko,lp2}. These extra channels are p-wave suppressed so they can contribute to thermal cross section at freezeout while being temperature-suppressed at present \cite{lp2}. Furthermore, the Higgs-singlet coupling can be constrained by DM direct detection and Higgs production cross section at the LHC \cite{lp2}.

On the other hand, in Model II, the vector-like doublet lepton couples to the axion but the vector-like singlet does not, so the $\gamma\gamma$ channel for DM annihilation has a smaller branching fraction than in Model I. For simplicity, in the maximal mixing case with $m_l=m_e$, $m_x={\tilde m}_x$, from the axion interactions in eqs.~(\ref{axionint1}) and (\ref{axionint2}), the DM annihilation cross section into a photon pair in Model II is reduced by a factor $4$ as compared to the one in Model I. Furthermore, there are extra diagrams producing $WW, Z\gamma$ from the DM annihilations due to the level-changing axion interactions, so the branching fraction of the $\gamma\gamma$ channel in Model II is reduced roughly by a factor $3$ to about $14\%$ as compared to Model I, as expected from the anomalies.
Therefore, the extra channels coming from the scalar partner of the axion should give a smaller contribution of at least $\langle\sigma v\rangle_X=(0.8-2.7)\langle\sigma v\rangle_a$ at freezeout than in Model I.

Finally, in Model III, instead of a vector-like singlet, a triplet is introduced with the Yukawa couplings to the axion and the Higgs. In this case, although the annihilation cross section into a photon pair is similar to half the one into $Z\gamma$, the two-photon line can still explain the Fermi gamma-ray line at $130\,{\rm GeV}$ \cite{tait}. The DM annihilation cross section into a photon pair is about $6\%$ in the limit of large extra lepton masses and it increases a bit when the lighter charged state is light.  Therefore, in Model III, without extra annihilation channels at freezeout, the axion mediation only can accommodate Fermi gamma-ray line by dark matter with thermal cross section.

\subsection{Electroweak precision constraints}

With the improved value of the observed W-boson mass \cite{wbosonmass}, the global electroweak precision analysis with the Higgs mass at $m_h=125\,{\rm GeV}$
gives rise to the constraint on the $STU$ parameters as \cite{mtrott}
\be
S=0.00\pm 0.10,\quad T=0.02\pm 0.11,\quad U=0.03\pm 0.09.
\ee
The contribution of the vector-like extra leptons to the $T$ parameter \cite{deltaT} is 
\bea
\Delta T&=& \frac{1}{16\pi s^2_W c^2_W} \bigg[ \sum_i (|V_{1i}|^2+|U_{1i}|^2) \theta_+(y_i,y_l)
+2{\rm Re}(V_{1i} U^*_{1i} )\theta_-(y_i, y_l)   \nonumber \\
&&-(|V_{12} V_{11}|^2 +|U_{12} U_{11}|^2) \theta_+(y_1,y_2)- 2{\rm Re}(V_{12}V^*_{11} U^*_{12}U_{11}) \theta_-(y_1,y_2) 
\bigg]
\eea
where $y_i\equiv m^2_i/m^2_Z$, $y_l\equiv m^2_l /m^2_Z$, and
\bea
\theta_+(a,b)&\equiv&a+b-\frac{2ab}{a-b}\,\ln\frac{a}{b},  \\
\theta_-(a,b)&\equiv& 2\sqrt{ab} \Big(\frac{a+b}{a-b}\,\ln\frac{a}{b}-2\Big).
\eea
In the case with $m_l=m_e$, $m_x={\tilde m}_x$, the above contribution to the $T$ parameter becomes
\bea
\Delta T&=& \frac{1}{16\pi s^2_W c^2_W} \bigg[\theta_+(y_1,y_l)+\theta_+(y_2,y_l)+\theta_-(y_1, y_l)+  \theta_-(y_2, y_l) \nonumber \\
&&-\frac{1}{2}\theta_+(y_1,y_2)- \frac{1}{2}\theta_-(y_1,y_2) 
\bigg].
\eea
As $\Delta T$ is more stringent than $\Delta S$ in the models \cite{wagner},  it is enough to impose  $\Delta T$ for the EWPD constraint in next section.

\subsection{Higgs to di-photon decay rate}

\begin{figure}[t]
\centering
\includegraphics[width=14cm]{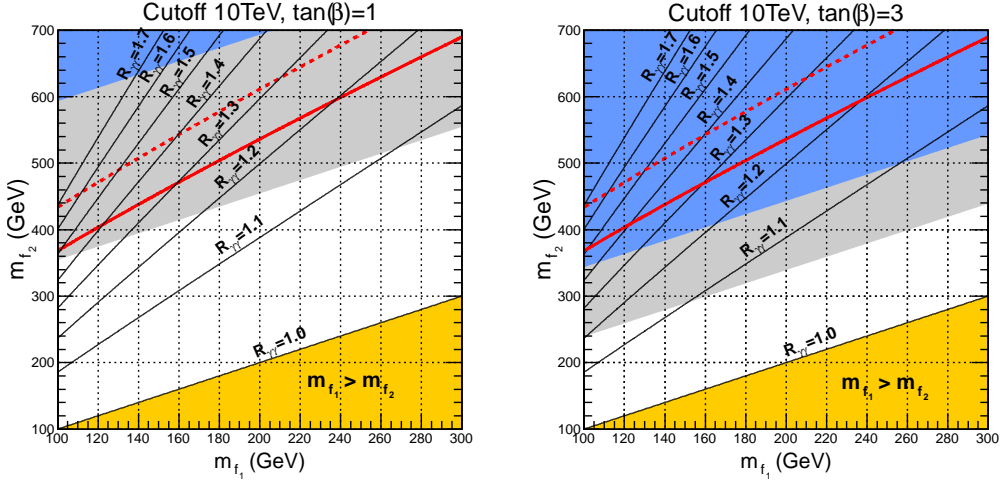} 
\caption{Parameter space of $m_{f_1}$ vs $m_{f_2}$ for the singlet models with cutoff $M_I=10\,{\rm TeV}$. Solid lines show $R_{\gamma\gamma}=1.0-1.7$  with step $0.1$ from bottom to top.  
The region above red solid(dashed) line is disfavored by EWPD at $68(95)\%$ C.L. Blue region is excluded by perturbativity while gray region is excluded by vacuum metastability. $\tan\beta=1(3)$ is chosen for the left(right) plot. We took ${\tilde y}_l=y_l/\tan\beta$ for $m_x={\tilde m}_x$ and $m_h=125\,{\rm GeV}$ for Higgs mass. }
\label{fig:rgg1}
\end{figure}

The heavy charged leptons, introduced to explain the Fermi gamma-ray line, can also lead to extra contributions to the decay of the SM-like Higgs into a photon pair through the Higgs couplings. 
In this section, we discuss the impact of EWPD constraints, perturbativity and vacuum stability conditions on the parameter space for which the Higgs diphoton signal is enhanced. 
Furthermore, we show how Fermi gamma-ray line restricts the parameter space further.

We define the ratio of the Higgs production cross section times the branching fraction, $\mu_{\gamma\gamma}\equiv \frac{\sigma\times {\rm Br}_{\gamma\gamma}}{(\sigma\times {\rm Br}_{\gamma\gamma})_{SM}}$.
The reported signal strengths for the Higgs-like events from ATLAS and CMS are the following \cite{july4},
\bea
\mu^{\rm ATLAS}_{\gamma\gamma} =1.90\pm 0.5, \quad
\mu^{\rm CMS}_{\gamma\gamma}=1.56\pm 0.43 ,\quad \mu^{\rm combi}_{\gamma\gamma}=1.71\pm 0.33.
\eea
The excess in the Higgs-to-diphoton signal might come from a statistical fluctuations or systematic uncertainties. Furthermore, it might well be that the QCD uncertainties were underestimated such that the deviations from the SM Higgs couplings could not be so significant \cite{djouadi}.
Nonetheless, we investigate how much the Higgs diphoton signal is enhanced by the Higgs couplings to extra vector-like leptons in our models.

\begin{figure}[t]
\centering
\includegraphics[width=14cm]{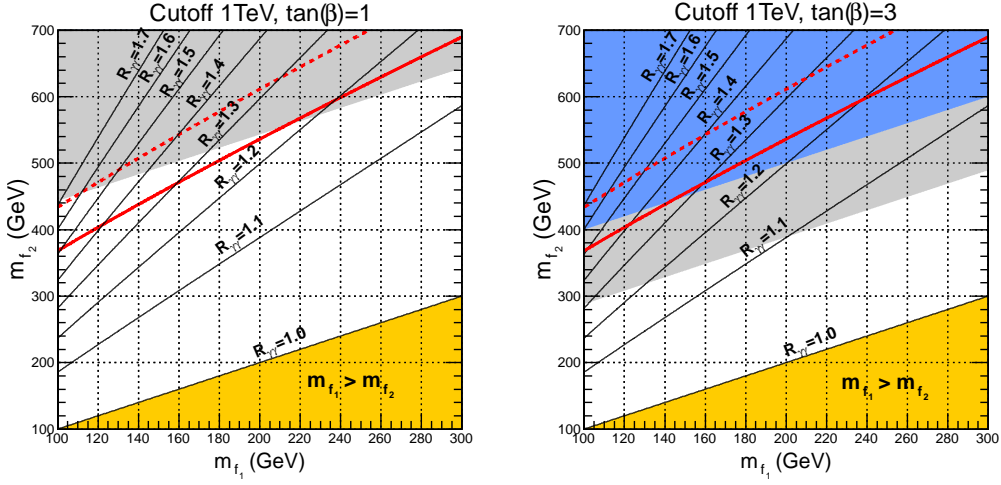} 
\caption{The same as in Fig.~\ref{fig:rgg1}, but with cutoff $M_I=1\,{\rm TeV}$. 
 }
\label{fig:rgg2}
\end{figure}

In the presence of the effective Higgs coupling to charged fermions with $c_f=\frac{M}{v} y_f $ in eq.~(\ref{effhiggs}),
the ratio of the Higgs to di-photon decay rate with respect to the SM value becomes
\be
R_{\gamma\gamma}=\bigg|1+y_f\frac{v}{m_f}\frac{A_f(\tau_f)}{A_V(\tau_W)+N_c Q^2_t A_f(\tau_t)}\bigg|^2
\ee
where $N_c=3$ is the number of color, $Q_t=+\frac{2}{3}$ is the top quark electric charge in units of $|e|$, and $\tau_i\equiv 4m^2_i/m^2_h$, $i=t,W,f$. Below the WW threshold the loop functions are given by
\bea
A_V(x)&=& -x^2 \Big[2x^{-2}+3 x^{-1}+3(2x^{-1}-1)f(x^{-1})\Big], \\
A_f(x)&=& 2x^2 \Big[x^{-1}+(x^{-1}-1) f(x^{-1})\Big]
\eea
with $f(x)={\rm arcsin}^2\sqrt{x}$ for $x\geq 1$. 
In the case with two charged Dirac fermions, 
the Higgs to di-photon decay rate is
\be
R_{\gamma\gamma}=\left|1+v\,\frac{y_{f_1}A_f(\tau_1)/m_{f_1}+y_{f_2}A_f(\tau_2)/m_{f_2}}{A_V(\tau_W)+N_c Q^2_t A_f(\tau_t)}\right|^2.
\ee
We note that the above ratio can be compared to the experimental measure by $R_{\gamma\gamma}\simeq \mu_{\gamma\gamma}$ when the singlet mixing to the SM Higgs boson is small enough.

\begin{figure}[t]
\centering%
\includegraphics[width=14cm]{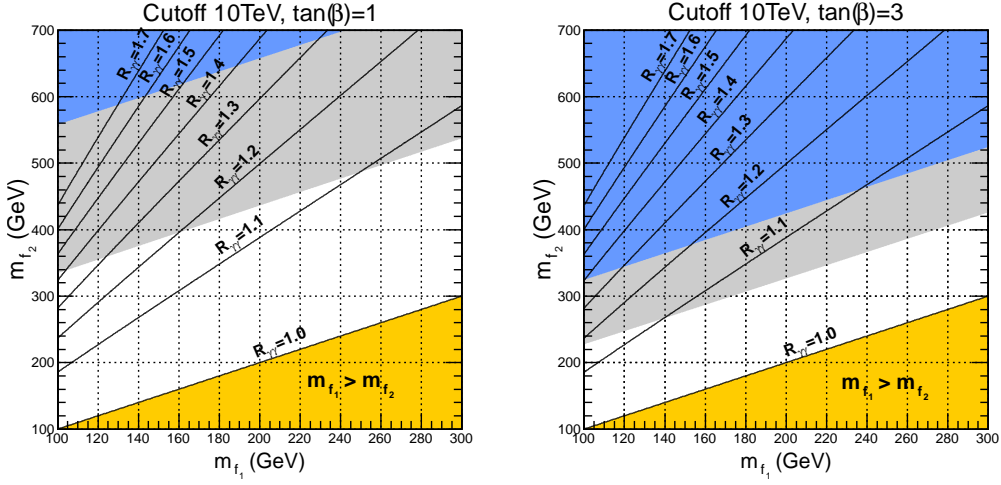} 
\caption{Parameter space of $m_{f_1}$ vs $m_{f_2}$ for the triplet model with cutoff $M_I=10\,{\rm TeV}$. Solid lines show $R_{\gamma\gamma}=1.0-1.7$  with step $0.1$ from bottom to top.  
Blue region is excluded by perturbativity while gray region is excluded by vacuum metastability. $\tan\beta=1(3)$ is chosen for the left(right) plot. We took ${\tilde y}_l=y_l/\tan\beta$ for $m_x={\tilde m}_x$ and $m_h=125\,{\rm GeV}$ for Higgs mass. }
\label{fig:rgg3}
\end{figure}

\begin{figure}[t]
\centering%
\includegraphics[width=14cm]{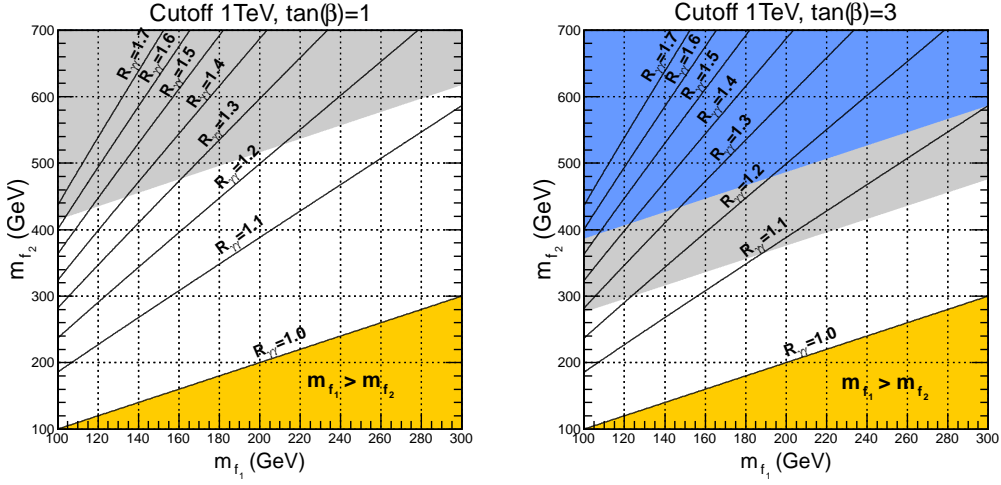} 
\caption{The same as in Fig.~\ref{fig:rgg3}, but with cutoff $M_I=1\,{\rm TeV}$. 
 }
\label{fig:rgg4}
\end{figure}

In Figs.~\ref{fig:rgg1} and \ref{fig:rgg2}, in the case of the maximal mixing of extra charged fermions for Model I and II (the singlet models), we show the parameter space giving rise to the Higgs to diphoton decay rate larger than the SM value, depending on the cutoff scale $M_I$ and $\tan\beta=v_u/v_d$. We have overlaid EWPD constraints at $68\%$ and $95\%$ C.L., perturbativity and vacuum metastability by using the renormalization group equations in appendix D.  We didn't include the coupling between the Higgs and the scalar partner of the axion in our analysis, although it could improve vacuum stability a bit if sizable.
Furthermore, we assume that the axion couplings to dark matter and extra leptons are perturbative at least until the vacuum stability scale of the SM sector.
As far as EWPD constraints at $95\%$ C. L. are concerned, the Higgs to di-photon decay rate can be enhanced up to $R_{\gamma\gamma}\simeq1.6$ for $m_{f_1}\lesssim115\,{\rm GeV}$ and $m_{f_2}\lesssim 450\,{\rm GeV}$. However, vacuum stability bound is the strongest constraint for the case with a large mass splitting so the Higgs to di-photon decay rate can be enhanced to at most $R_{\gamma\gamma}\simeq 1.4$ at $m_{f_1}\lesssim 110\,{\rm GeV}$ and  $m_{f_2}\lesssim 450\,{\rm GeV}$ for $M_I=10\,{\rm TeV}$ and $R_{\gamma\gamma}\simeq 1.6$ at $m_{f_1}\lesssim 110\,{\rm GeV}$ and $m_{f_2}\lesssim 360\,{\rm GeV}$ for $M_I=1\,{\rm TeV}$.  Consequently, in order to get the Higgs to diphoton rate to be compatible with the LHC within $1\sigma$, we need a lighter charged extra lepton to be around $110\,{\rm GeV}$. 

In Figs.~\ref{fig:rgg3} and \ref{fig:rgg4}, we show the parameter space for Model III (the triplet model) with perturbativity and vacuum metastability. Due to the more stringent vacuum stability bounds, we get 
at most $R_{\gamma\gamma}\simeq 1.3$ at $m_{f_1}\lesssim 120\,{\rm GeV}$ and  $m_{f_2}\lesssim 360\,{\rm GeV}$ for $M_I=10\,{\rm TeV}$ and $R_{\gamma\gamma}\simeq 1.5$ at $m_{f_1}\lesssim 115\,{\rm GeV}$ and $m_{f_2}\lesssim 440\,{\rm GeV}$ for $M_I=1\,{\rm TeV}$. 

\begin{table}[h!]\small
\begin{center}
\begin{tabular}{c||c|c|c}
 \hline
 & Model I & Model II & Model III
\\\hline\hline
$(c_1,c_2)$ & $(3,1)$ & $(1,1)$ & $(1,3)$
\\
${\rm Br}({\bar\chi}\chi \rightarrow \gamma\gamma)$ & $\gtrsim 40\%$  &  $\gtrsim 14\%$   & $\gtrsim  6\%$
\\
$R_{\gamma\gamma}$ &  $ \lesssim 1.5$  &   $\lesssim 1.5$    & $\lesssim 1.4$ 
\\ 
\hline
\end{tabular}
\caption{Summary of the model predictions}
\label{table:summary}
\end{center}
\end{table}

From the Fermi gamma-ray line, we require the tree-level DM annihilation into a pair of lighter charged leptons not to contribute, because otherwise the tree-level channel would easily dominate the DM annihilation into a photon pair.
Then, we need the lighter charged extra lepton to be larger than dark matter mass, $m_{\chi}=130\,{\rm GeV}$, in both singlet and triplet models.  Therefore, for $m_{f_1}>130\,{\rm GeV}$, in combination with vacuum stability conditions, the Higgs-to-diphoton rate in the singlet(triplet) model is bounded by $R_{\gamma \gamma} <1.4(1.3)$ for the cutoff at $M_I=10\,{\rm TeV}$ and $R_{\gamma \gamma} <1.5(1.4)$ for the cutoff at $M_I=1\,{\rm TeV}$.
The model predictions for the DM annihilation cross section into a pair and the Higgs-to-diphoton decay rate are summarized in Table~\ref{table:summary}.

\subsection{Cosmology and collider bounds on extra leptons}

If extra charged leptons do not couple to the SM leptons directly as is the case in our minimal models, the extra  lighter charged lepton would be stable. Then, it would be dangerous for the successful Big Bang nucleosynthesis(BBN) and the collider limits on stable charged particles. 

First, from the success of the BBN, we could put a bound on the lifetime of the lighter charged lepton to be smaller than about $5\times 10^3\, {\rm sec}$, similarly to the case of stay NLSP in the MSSM \cite{BBN}, by considering the bound state effects of charged particle during BBN epoch. 
Furthermore, if the lighter charged lepton is stable,  there are strong bounds on the mass of stable charged states at the LHC \cite{colliders} so that one could not get a large Higgs to diphoton decay rate. Therefore, we need to extend the models \cite{wagner,AH}. First, we can introduce a mixing of the extra doublet neutrino with an SM singlet neutrino. In this case, the lighter charged fermion can decay into $W$-boson and the SM singlet neutrino sufficiently fast to satisfy the BBN and collider bounds.
Second, extra charged lepton can mix with the SM charged leptons by the renormalizable Yukawa couplings such that the lighter charged lepton decays by $f_1
\rightarrow Zl$ and $f_1\rightarrow W \nu$.
 In this case, however, one has to take the sufficiently small mixing terms to satisfy the constraints coming from flavor violation such as $\mu\rightarrow e\gamma$ and $\tau\rightarrow \mu\gamma$ and the $Z$-width. Requiring the mixing angle $|U_{il_4}|\lesssim 10^{-2}$  \cite{AH}, we can satisfy those constraints and make the lighter charged lepton decay prompt.

\section{Conclusions}

We have considered the possibility that extra vector-like leptons enhance the Higgs-to-diphoton rate and generate anomaly interactions in axion-mediated dark matter models. In order to get a large DM annihilation cross section to monochromatic photons as hinted from the Fermi gamma-ray line, the tree-level channels for DM annihilations must be suppressed or forbidden, constraining the dark matter models. In our models, the masses of the extra leptons that couple to the axion mediator must be larger than dark matter mass. 
Furthermore, dark matter with thermal cross section restricts the branching fraction of the DM annihilation cross section into a photon pair, favoring the triplet model in which the $SU(2)_L$ anomalies are larger than in the singlet models. 

The enhancement of the Higgs-to-diphoton rate requires new large couplings between the Higgs and new light charged states, so it can be comparable to the signal strength observed at the LHC at the expense of introducing a tension with EWPD and/or a low cutoff due to perturbativity and vacuum stability. In the models with extra vector-like leptons, Fermi gamma-ray line can give new bounds on the masses of the extra leptons, which constrain the region of the enhanced Higgs-to-diphoton rate. When Fermi gamma-ray line is explained by the two-photon line, which is the case in all the minimal models, the lighter charged lepton must be heavier than $130\,{\rm GeV}$, which in turn bounds the ratio of Higgs-to-diphoton rate to the SM value to at most $1.4(1.5)$ even for the low cutoff of $10(1)\,{\rm TeV}$.

\section*{Acknowledgments}

MP is partially supported by a CERN-Korean fellowship.

\def\theequation{A.\arabic{equation}}

\setcounter{equation}{0}

\vskip0.8cm
\noindent
{\Large \bf Appendix A: Axion couplings}
\vskip0.4cm
\noindent

From eq.~(\ref{leptonY}) with eqs.~(\ref{rot1}) and (\ref{rot2}),  in the basis of mass eigenstates, we obtain the interaction terms for the lighter axion and the lighter CP-even Higgs in terms of Weyl spinors as
\bea
-{\cal L}_{\rm int}
&=& i\frac{m_\chi}{v_s}\,a \chi {\tilde \chi}+ i\frac{m_l}{v_s}\, a\nu_4 {\tilde \nu}_4 
+i a \,(\lambda_{1}  f_1 {\tilde f}_1+
\lambda_{2}  f_2 {\tilde f}_2 +\lambda_{3}f_1 {\tilde f}_2 +\lambda_{4} f_2 {\tilde f}_1) \nonumber \\
&&- h \,(y_{1}  f_1 {\tilde f}_1+
y_{2}  f_2 {\tilde f}_2 +y_{3}f_1 {\tilde f}_2 +y_{4} f_2 {\tilde f}_1) +{\rm h.c.}
 \label{yukawas0}
\eea
where $\alpha$ is the mixing angle between two CP-even Higgs bosons and the Higgs couplings for all models that we considered in the text are
\bea
y_{1}&=&\frac{\sin\alpha}{v_d m_{f_1}}\Big(m^2_x \cos^2\theta_f -\frac{1}{2}m_e m_x\sin 2\theta_f\Big)
-\frac{\cos\alpha}{v_u m_{f_1}}\Big({\tilde m}^2_x \sin^2\theta_f-\frac{1}{2}m_l {\tilde m}_x\sin 2\theta_f \Big), ~~~~ ~~\\
y_{2}&=&\frac{\sin\alpha}{v_d m_{f_2}}\Big(m^2_x \sin^2\theta_f +\frac{1}{2}m_e m_x\sin 2\theta_f\Big)-\frac{\cos\alpha}{v_u m_{f_2}}\Big({\tilde m}^2_x\cos^2\theta_f+\frac{1}{2}m_l {\tilde m}_x\sin 2\theta_f\Big), \\
y_{3}&=& \frac{\sin\alpha}{v_d m_{f_2}}\Big(m_e m_x\cos^2\theta_f+\frac{1}{2}m^2_x\sin 2\theta_f\Big)+\frac{\cos\alpha}{v_u m_{f_2}}\Big(m_l {\tilde m}_x \sin^2\theta_f+\frac{1}{2} {\tilde m}^2_x\sin2\theta_f \Big), \\
y_{4}&=& \frac{\sin\alpha}{v_d m_{f_1}}\Big(\frac{1}{2}m^2_x\sin 2\theta_f-m_e m_x\sin^2\theta_f\Big)-\frac{\cos\alpha}{v_u m_{f_1}}\Big(m_l {\tilde m}_x \cos^2\theta_f-\frac{1}{2} {\tilde m}^2_x\sin2\theta_f \Big),
\eea
and the axion couplings are: for Model I,
\bea
\lambda_{1}&=&\frac{1}{v_s m_{f_1}}\Big(m^2_l \cos^2\theta_f +m^2_e \sin^2\theta_f-\frac{1}{2}(m_l {\tilde m}_x+m_e m_x)\sin2\theta_f\Big),  \\
\lambda_{2}&=&\frac{1}{v_s m_{f_2}}\Big(m^2_l \sin^2\theta_f +m^2_e \cos^2\theta_f+\frac{1}{2}(m_l {\tilde m}_x+ m_e m_x)\sin2\theta_f \Big), \\
\lambda_{3}&=& \frac{1}{v_s m_{f_2}}\Big(m_l {\tilde m}_x\cos^2\theta_f-m_e m_x \sin^2\theta_f +\frac{1}{2}(m^2_l-m^2_e) \sin2\theta_f \Big), \\
\lambda_{4}&=& \frac{1}{v_s m_{f_1}}\Big(m_e m_x \cos^2\theta_f-m_l {\tilde m}_x\sin^2\theta_f+\frac{1}{2}(m^2_l-m^2_e)\sin2\theta_f \Big);
\eea
for Model II,
\bea
\lambda_{1}&=&\frac{m_l \cos\theta_f}{v_s m_{f_1}}(m_l\cos\theta_f-{\tilde m}_x \sin\theta_f),  \\
\lambda_{2}&=&\frac{m_l\sin\theta_f}{v_s m_{f_2}}(m_l\sin\theta_f+{\tilde m}_x\cos\theta_f), \\
\lambda_{3}&=& \frac{m_l \cos\theta_f}{v_s m_{f_2}}(m_l\sin\theta_f+{\tilde m}_x\cos\theta_f), \\
\lambda_{4}&=& \frac{m_l \sin\theta_f}{v_s m_{f_1}}(m_l\cos\theta_f-{\tilde m}_f \sin\theta_f).
\eea
In the decoupling limit of the extra Higgs fields, we need to impose $\alpha=\beta-\frac{\pi}{2}$ so that we get $-\sin\alpha/v_d=\cos\alpha/v_u=1/v$.

\def\theequation{B.\arabic{equation}}

\setcounter{equation}{0}

\vskip0.8cm
\noindent
{\Large \bf Appendix B:  Gauge couplings to extra leptons}
\vskip0.4cm
\noindent

For a pair of left-handed lepton doublet $(\nu_4,e_4)^T$ and right-handed singlet $e^c_4$, the electroweak interactions are
\bea
{\cal L}_{\rm gauge}&=& \frac{g}{2\sqrt{2}} W^\dagger_\mu 
\Big({\bar \nu}_4 \gamma^\mu (1-\gamma^5)e +{\rm h.c.} ) \nonumber \\
&&+\frac{e}{2\sin\theta_W\cos\theta_W}\,
Z_\mu\sum_{\nu_4,e} {\bar f}\gamma^\mu (v_f -a_f \gamma^5 ) f \nonumber \\
&&-e A_\mu {\bar e}\gamma^\mu e
\eea
where $e\equiv(e_4,e^{c\dagger}_4)^T$,
and
\bea
v_{\nu_4}&=&a_{\nu_4}=\frac{1}{2}, \\
v_e&=&-\frac{1}{2}+2\sin^2\theta_W, \quad a_e=-\frac{1}{2}.
\eea
Thus, for a vector-like lepton doublet and a vector-like lepton singlet, we need to include the same electroweak interactions for the vector-like partners, a right-handed lepton doublet $({\tilde \nu}^\dagger_4, {\tilde e}^\dagger_4)$ and a left-handed singlet ${\tilde e}^{c\dagger}_4$. 
Then, we get the electroweak interactions for them in terms of Dirac spinors as
\bea
{\cal L}_{\rm gauge}&=& \frac{g}{2\sqrt{2}} W^\dagger_\mu 
\Big({\bar \nu} \gamma^\mu (1-\gamma^5)e
+{\bar \nu} \gamma^\mu (1+\gamma^5)e' +{\rm h.c.} \Big) \nonumber \\
&&+\frac{e}{2\sin\theta_W\cos\theta_W}\,
Z_\mu\Big({\bar \nu} \gamma^
\mu\nu+\sum_{e,e'} {\bar f}\gamma^\mu (v_f -a_f \gamma^5 ) f \Big)\nonumber \\
&&-e A_\mu ({\bar e}\gamma^\mu e+{\bar e}'\gamma^\mu  e' )
\eea
where $\nu\equiv(\nu_4,{\tilde\nu}^\dagger_4)^T$, $e'\equiv({\tilde e}^c_4,{\tilde e}^\dagger_4)^T$, and the neutral current interactions are $v_{e'}=v_e$ and $a_e=-a_{e'}=-\frac{1}{2}$.
Therefore, using the rotation matrices in eqs.~(\ref{rot1}) and (\ref{rot2}),  the above interactions become in the basis of mass eigenstates,
\bea
{\cal L}_{\rm gauge}&=& \frac{g}{2\sqrt{2}} W^\dagger_\mu 
\Big({\bar \nu} \gamma^\mu (1-\gamma^5)V_{1i} F_i
+{\bar \nu} \gamma^\mu (1+\gamma^5)U_{1i}F_i+{\rm h.c.} \Big) \nonumber \\
&&+\frac{e}{2\sin\theta_W\cos\theta_W}\,
Z_\mu\Big[{\bar \nu} \gamma^\mu\nu+v_e ({\bar F}_1\gamma^\mu F_1 +{\bar F}_2 \gamma^\mu F_2)\nonumber \\
&&\quad+\frac{1}{2}a_e (V^\dagger_{j1} V_{1i}-V^\dagger_{j2} V_{2i} ){\bar F}_j \gamma^\mu (1-\gamma^5)F_i
+ \frac{1}{2}a_e (U^\dagger_{j1} U_{1i}-U^\dagger_{j2} U_{2i} ){\bar F}_j \gamma^\mu (1+\gamma^5)F_i \Big] \nonumber \\
&&-A_\mu ({\bar F}_1\gamma^\mu F_1+{\bar F}_2\gamma^\mu  F_2 )
\eea
where $F_1\equiv(f_1,{\tilde f}^\dagger_1)^T$ and $F_2\equiv(f_2,{\tilde f}^\dagger_2)^T$ are Dirac spinors with masses $m_{f_1}, m_{f_2}$.

\def\theequation{C.\arabic{equation}}

\setcounter{equation}{0}

\vskip0.8cm
\noindent
{\Large \bf Appendix C: Matrix elements for the axion decay}
\vskip0.4cm
\noindent

In this section, we present the expressions for the matrix elements for the axion decay.

Suppose that the vector-like fermions, $\psi_i=(f_1,f_2,\nu)$, have the following axion and gauge interactions,
\be
{\cal L}_f=i \lambda_{\psi_{ij}} a {\bar \psi}_i \gamma^5 \psi_j +{\bar \psi}_i \gamma^\mu(g_{V_{ij}} -g'_{V_{ij}}\gamma^5) \psi_j V_\mu.
\ee
We consider the case that $g'_{V_{ij}}=0$ and $\lambda_{\psi_{ij}}$, $g_{W_{ij}}$ and $g_{\gamma_{ij}}$ are diagonal as for the simple charged fermion mass matrix with $m_l=m_e$ and $m_x={\tilde m}_x$ in Model I. So, we denote the diagonal components by $\lambda_{\psi_i}\equiv\lambda_{\psi_{ii}}$, $g_{V_i}\equiv g_{V_{ii}}$ and $g_{\gamma_i}\equiv g_{\gamma_{ii}}$. Then, the matrix elements for the decay channels of the axion are ${\cal M}=\epsilon^{*\mu}(k_1)\epsilon^{*\nu}(k_2){\cal M}_{\mu\nu}$, with
\bea
{\cal M}_{\mu\nu}(a\rightarrow \gamma\gamma)=\frac{i}{4\pi^2}\epsilon_{\mu\nu\rho\sigma} k^\rho_2 k^\sigma_1\sum_i \frac{\lambda_{f_i}g^2_{\gamma i}}{ m_{f_i}} \, A_1(\tau_i),
\eea
\bea
{\cal M}_{\mu\nu}(a\rightarrow ZZ) &=& 
\frac{i}{4\pi^2}\epsilon_{\mu\nu\rho\sigma} k^\rho_2 k^\sigma_1\,\bigg[\sum_i  \frac{\lambda_{f_i}g^2_{Z _i}}{ m_{f_i}} A_2(\tau_i,\rho^Z_i)
 +\frac{\lambda_{\nu}g^2_{Z_\nu}}{ m_\nu} \, A_2(\tau_\nu,\rho^Z_\nu) \nonumber \\
&&+\frac{\lambda_{f_1}g^2_{Z_{12}}}{ m_{f_1}} \, A_3(\tau_1,\rho^Z_1,\frac{m_{f_2}}{m_{f_1}})  
+\frac{\lambda_{f_2}g^2_{Z_{12}}}{ m_{f_2}}\, A_3(\tau_2,\rho^Z_2,\frac{m_{f_1}}{m_{f_2}})\bigg], ~~
\eea
\bea
{\cal M}_{\mu\nu}(a\rightarrow Z\gamma)=
\frac{i}{4\pi^2}\epsilon_{\mu\nu\rho\sigma} k^\rho_2 k^\sigma_1\sum_i  \frac{\lambda_{f_i}g_{Z _i} g_{\gamma_i}}{ m_{f_i}} \, A_4(\tau_i,\rho^Z_i), 
\eea
\be
{\cal M}_{\mu\nu}(a\rightarrow WW)=   \frac{i}{4\pi^2}\epsilon_{\mu\nu\rho\sigma} k^\rho_2 k^\sigma_1\bigg[
\sum_i  \frac{\lambda_{\nu}g^2_{W _{i\nu}}}{ 2m_\nu} \, A_3(\tau_\nu,\rho^W_\nu,\frac{m_i }{m_\nu}) 
+\sum_i  \frac{\lambda_{f_i}g^2_{W _{i\nu}}}{2 m_{f_i}} \, A_3(\tau_i,\rho^W_i,\frac{m_\nu}{m_{f_i}})  \bigg]~~~~~
\ee
where $\tau_{1,2}\equiv 4m^2_{f_{1,2}}/m^2_a$, $\tau_\nu\equiv 4m^2_\nu /m^2_a$, and $\rho^V_{1,2} \equiv 4 m_{f_{1,2}}^2/m_V^2$, 
$\rho^V_{\nu} \equiv 4 m_{\nu}^2/m_V^2$ .
\bea
A_1(\tau)&=&2\int_0^1 dx \int^{1-x}_0 dy \,\frac{1}{\Delta_1(\tau)} = \tau\arcsin^2(1/\sqrt{\tau}), \\
A_2(\tau,\rho^V)&=& 2\int_0^1 dx \int^{1-x}_0 dy \, \frac{1}{\Delta_2(\tau,\rho^V)}, \\
A_3(\tau,\rho^V,b)&=&2 \int_0^1 dx \int^{1-x}_0 dy \, \frac{b+(1-b) (x+y)}{\Delta_3(\tau,\rho^V,b)}, \\
A_4(\tau,\rho^V) &=&\int_0^1 dx \int^{1-x}_0 dy\, \Big(\frac{1}{\Delta_4(\tau,\rho^V)}+ \frac{1}{\Delta'_4(\tau,\rho^V)}\Big)  \\
&= &\frac{\tau \rho^V}{\tau -\rho^V}\left\{\arcsin^2\left(1/\sqrt{\rho^V}\right)-\arcsin^2\left(1/\sqrt{\tau}\right)\right\},
\eea
with
\bea
\Delta_1(\tau)&=&1- \frac{4}{\tau}\,x y, \\
\Delta_2(\tau,\rho^V)&=&\Delta_1(\tau)- \frac{4}{\rho^V}\,(x+y) (1-x-y), \\
\Delta_3(\tau,\rho^V,b)&=&\Delta_1(\tau)- \frac{4}{\rho^V}\, (x+y)(1-x-y)- (1-b^2) (1-x-y), \\
\Delta_4(\tau,\rho^V)&=&\Delta_1(\tau)- \frac{4}{\rho^V}\, y (1-x-y), \\
\Delta'_4(\tau,\rho^V)&=& \Delta_1(\tau)- \frac{4}{\rho^V}\, x (1-x-y).
\eea

Then, the axion decay rates are
\bea
\Gamma_a(a\rightarrow\gamma\gamma)&=& \frac{1}{32\pi m_a}|{\cal M}_{a\rightarrow \gamma\gamma}|^2 \nonumber  \\
&=&\frac{m^3_a}{64\pi^3}\bigg|\sum_i \frac{\lambda_{f_i}\alpha_{\gamma i}}{ m_{f_i}} \, A_1(\tau_i)\bigg|^2,\\
\Gamma_a(a\rightarrow ZZ)&=&\frac{1}{32\pi m_a}|{\cal M}_{a\rightarrow ZZ}|^2 \nonumber  \\
&=&\frac{m^3_a}{64\pi^3}\Big(1-\frac{4m^2_Z}{m^2_a}\Big)^{3/2}
\bigg|\sum_i  \frac{\lambda_{f_i}\alpha_{Z _i}}{ m_{f_i}} A_2(\tau_i,\rho^Z_i) 
+\frac{\lambda_{\nu}\alpha_{Z_\nu}}{ m_{\nu}} \, A_2(\tau_\nu,\rho^Z_\nu)\nonumber \\
&&+ \frac{\lambda_{f_1}\alpha_{Z_{12}}}{ m_{f_1}} \, A_3(\tau_1,\rho^Z_1,\frac{m_{f_2}}{m_{f_1}}) 
 +\frac{\lambda_{f_2}\alpha_{Z_{12}}}{ m_{f_2}}\, A_3(\tau_2,\rho^Z_2,\frac{m_{f_1}}{m_{f_2}}) 
 \bigg|^2,~~~ \\
\Gamma_a(a\rightarrow Z\gamma)&=&\frac{1}{16\pi m_a}|{\cal M}_{a\rightarrow Z\gamma}|^2 \nonumber  \\
&=&\frac{m^3_a}{32\pi^3}\Big(1-\frac{m^2_Z}{m^2_a}\Big)^3
\bigg|\sum_i  \frac{\lambda_{f_i}\sqrt{\alpha_{Z _i} \alpha_{\gamma_i}}}{ m_{f_i}} \, A_4(\tau_i,\rho^Z_i)\bigg|^2, \\
\Gamma_a(a\rightarrow WW)&=&\frac{1}{16\pi m_a}|{\cal M}_{a\rightarrow WW}|^2 \nonumber \\
&=&\frac{m^3_a}{32\pi^3}\Big(1-\frac{4m^2_W}{m^2_a}\Big)^{3/2}
\bigg|\sum_i  \frac{\lambda_\nu\alpha_{W _{i\nu}}}{ 2m_\nu} \, A_3(\tau_\nu,\rho^W_\nu,\frac{m_i }{m_\nu}) \nonumber \\
&&+\sum_i  \frac{\lambda_{f_i}\alpha_{W _{i\nu}}}{ 2m_{f_i}} \, A_3(\tau_i,\rho^W_i,\frac{m_\nu}{m_{f_i}})  \bigg|^2\, ,
\eea 
with $\alpha_{V_i}\equiv g^2_{V_i}/4\pi$, etc.
When the axion is kinematically allowed to decay into a dark matter pair, that is, for $m_a > 2m_\chi$, 
we need to include the extra channel with decay rate given by
\be
\Gamma_a(a\rightarrow {\bar\chi}\chi)=\frac{|\lambda_\chi|^2}{16\pi}\, m_a \Big(1-\frac{4m^2_\chi}{m^2_a}\Big)^{1/2}.
\ee

Furthermore, the amplitudes for the DM annihilation into a pair of gauge bosons, $V_1$ and $V_2$, are
\be
{\cal M}_{\chi {\bar \chi}\rightarrow\gamma\gamma}=({\cal M}_{\chi{\bar\chi}\rightarrow a})\,\left(\frac{i}{s-m^2_a-i\Gamma_a m_a}\right)\,({\cal M}_{a\rightarrow V_1V_2})\,,
\ee
where $m^2_a$, $\tau_i$ in ${\cal M}_{\chi{\bar\chi}\rightarrow a}$ and ${\cal M}_{a\rightarrow V_1V_2}$ are replaced by $s$ and $4m^2_i/s$, respectively, in the axion decay amplitude part. 
As a consequence, the annihilation cross sections are given by
\bea
\langle\sigma v\rangle_{\gamma\gamma}&=&\frac{1}{16\pi s}|{\cal M}_{\chi{\bar\chi}\rightarrow \gamma\gamma}|^2 \nonumber  \\
&=&\frac{|\lambda_\chi|^2}{512\pi^3}\frac{s^2}{(s-m^2_a)^2+\Gamma^2_a m^2_a}
\left| \sum_i \frac{\lambda_{f_i}\alpha_{\gamma i}}{ m_{f_i}} \, A_1(\tau_i)\right|^2, \\
\langle\sigma v\rangle_{ZZ}&=&\frac{1}{16\pi s}|{\cal M}_{\chi{\bar\chi}\rightarrow ZZ}|^2 \nonumber \\
&=&\frac{|\lambda_\chi|^2}{512\pi^3}\frac{s^2}{(s-m^2_a)^2+\Gamma^2_a m^2_a}
\Big(1-\frac{m^2_Z}{m^2_\chi}\Big)^{3/2}\bigg|\sum_i  \frac{\lambda_{f_i}\alpha_{Z _i}}{ m_{f_i}} A_2(\tau_i,\rho^Z_i)
+\frac{\lambda_{\nu}\alpha_{Z_\nu}}{ m_{\nu}} \, A_2(\tau_\nu,\rho^Z_\nu) \nonumber \\
&&+\frac{\lambda_{f_1}\alpha_{Z_{12}}}{ m_{f_1}} \, A_3(\tau_1,\rho^Z_1,m_{f_2}/m_{f_1})
+\frac{\lambda_{f_2}\alpha_{Z_{12}}}{ m_{f_2}}\, A_3(\tau_2,\rho^Z_2,m_{f_1}/m_{f_2})  \bigg|^2, \\
\langle\sigma v\rangle_{Z\gamma}&=&\frac{1}{8\pi s}|{\cal M}_{\chi{\bar\chi}\rightarrow Z\gamma}|^2 \nonumber  \\
&=&\frac{|\lambda_\chi|^2}{256\pi^3}\frac{s^2}{(s-m^2_a)^2+\Gamma^2_a m^2_a}
\Big(1-\frac{m^2_Z}{4m^2_\chi}\Big)^3\bigg|\sum_i  \frac{\lambda_{f_i}\sqrt{\alpha_{Z _i} \alpha_{\gamma_i}}}{ m_{f_i}} \, A_4(\tau_i,\rho^Z_i)\bigg|^2, \\
\langle\sigma v\rangle_{WW}&=&\frac{1}{8\pi s}|{\cal M}_{\chi{\bar\chi}\rightarrow WW}|^2 \nonumber \\
&=&\frac{|\lambda_\chi|^2}{256\pi^3}\frac{s^2}{(s-m^2_a)^2+\Gamma^2_a m^2_a}\Big(1-\frac{m^2_W}{m^2_\chi}\Big)^{3/2}
\bigg|\sum_i  \frac{\lambda_\nu\alpha_{W _{i\nu}}}{ 2m_\nu} \, A_3(\tau_\nu,\rho^W_\nu,m_i /m_\nu) \nonumber \\
&&+\sum_i  \frac{\lambda_{f_i}\alpha_{W _{i\nu}}}{2 m_{f_i}} \, A_3(\tau_i,\rho^W_i,m_\nu/m_{f_i})  \bigg|^2 .
\eea 


\def\theequation{D.\arabic{equation}}

\setcounter{equation}{0}

\vskip0.8cm
\noindent
{\Large \bf Appendix D: Renormalization group equations with extra leptons}
\vskip0.4cm
\noindent

In the presence of the Yukawa coupling of extra vector-like leptons to the Higgs, we consider the renormalization group (RG) equations below the masses of heavy Higgs states in the decoupling limit.

Following the results \cite{ramond},  we obtain the RGEs, $\frac{d p_i}{dt}=\beta_{p_i}$, with $t\equiv \ln(\mu/m_t)$,  with a vector-like doublet fermion and a vector-like singlet fermion, for the SM gauge couplings, the top Yukawa coupling,  the extra Yukawa couplings and the Higgs quartic coupling,
\bea
16\pi^2 \beta_{g'}=   \frac{53}{6} g^{'3},\quad
16\pi^2 \beta_{g} =   -\frac{15}{6} g^3,\quad
16 \pi^2 \beta_{g_3} = -7 g^3_3,
\eea
\bea
16\pi^2 \beta_{y_t}&=& y_t \Big(\frac{9}{2}y^2_t+y^2_l +{\tilde y}^2_l-\frac{17}{12}g^{\prime 2}-\frac{9}{4}g^2-8 g^2_3\Big), \\
16\pi^2 \beta_{y_l}&=& y_l \Big(\frac{5}{2}y^2_l+{\tilde y}^2_l+3y^2_t-\frac{15}{4}g^{\prime 2}-\frac{9}{4}g^2\Big), \\
16\pi^2 \beta_{{\tilde y}_l}&=& {\tilde y}_l \Big(\frac{5}{2}{\tilde y}^2_l+y^2_l+3 y^2_t-\frac{15}{4}g^{\prime 2}-\frac{9}{4}g^2\Big),
\eea
\bea
16\pi^2 \beta_{\lambda}&=& 24\lambda^2 -\Big(3 g^{\prime 2}+9 g^2\Big) \lambda+\frac{3}{8}\Big(g^{\prime 4}+2  g^{\prime 2}g^2+3g^4\Big)
 \nonumber \\
 &&+4(3 y^2_t+ y^2_l +{\tilde y}^2_l )\lambda - 2(3y^4_t +y^4_l+{\tilde y}^4_l ). 
\eea
We note that we have ignored the quartic couplings of the singlet mediator $S$ in the above RGEs. The Higgs-singlet coupling can contribute a positive term to the RG equation for the Higgs quartic coupling so it can help improve the vacuum stability a bit.

On the other hand, for a vector-like doublet fermion and a triplet fermion, we get the corresponding RGEs,
\bea
16\pi^2 \beta_{g'}=   \frac{45}{6} g^{'3},\quad
16\pi^2 \beta_{g} =   -\frac{7}{6} g^3,\quad
16 \pi^2 \beta_{g_3} = -7 g^3_3,
\eea
\bea
16\pi^2 \beta_{y_t}&=& y_t \Big(\frac{9}{2}y^2_t+\frac{3}{2}y^2_l +\frac{3}{2}{\tilde y}^2_l-\frac{17}{12}g^{\prime 2}-\frac{9}{4}g^2-8 g^2_3\Big), \\
16\pi^2 \beta_{y_l}&=& y_l \Big(\frac{11}{4}y^2_l+{\tilde y}^2_l+3 y^2_t-\frac{3}{4}g^{\prime 2}-\frac{33}{4}g^2\Big), \\
16\pi^2 \beta_{{\tilde y}_l}&=& {\tilde y}_l \Big(\frac{11}{4}{\tilde y}^2_l+y^2_l+3y^2_t -\frac{3}{4}g^{\prime 2}-\frac{33}{4}g^2\Big),
\eea
\bea
16\pi^2 \beta_{\lambda}&=& 24\lambda^2 -\Big(3 g^{\prime 2}+9 g^2\Big) \lambda+\frac{3}{8}\Big(g^{\prime 4}+2  g^{\prime 2}g^2+3g^4\Big)
 \nonumber \\
 &&+2(6 y^2_t+ 3y^2_l +3{\tilde y}^2_l )\lambda - \Big(6y^4_t +\frac{5}{2}y^4_l+\frac{5}{2}{\tilde y}^4_l \Big). 
\eea


\small



\begin{thebibliography}{999}


\bibitem{july4}
  G.~Aad {\it et al.}  [ATLAS Collaboration],
  Phys.\ Lett.\ B
  [arXiv:1207.7214 [hep-ex]];
  S.~Chatrchyan {\it et al.}  [CMS Collaboration],
  Phys.\ Lett.\ B
  [arXiv:1207.7235 [hep-ex]].






\bibitem{fermilat}
  W.~B.~Atwood {\it et al.}  [LAT Collaboration],
  Astrophys.\ J.\  {\bf 697} (2009) 1071
  [arXiv:0902.1089 [astro-ph.IM]].





\bibitem{fermilat2}
  A.~A.~Abdo {\it et al.},
  Phys.\ Rev.\ Lett.\  {\bf 104} (2010) 091302
  [arXiv:1001.4836 [astro-ph.HE]].


\bibitem{fermilat3}
  M.~Ackermann {\it et al.}  [Fermi-LAT Collaboration],
  Phys.\ Rev.\ Lett.\  {\bf 107} (2011) 241302
  [arXiv:1108.3546 [astro-ph.HE]].



\bibitem{fermilat4}
  F.~M.~Ackermann {\it et al.}  [LAT Collaboration],
  arXiv:1205.2739 [astro-ph.HE].









\bibitem{weniger}
  C.~Weniger,
  arXiv:1204.2797 [hep-ph].

\bibitem{ibarra}
  T.~Bringmann, X.~Huang, A.~Ibarra, S.~Vogl and C.~Weniger,
  arXiv:1203.1312 [hep-ph].





\bibitem{raidal}
  E.~Tempel, A.~Hektor and M.~Raidal,
  arXiv:1205.1045 [hep-ph].


\bibitem{su}
  M.~Su and D.~P.~Finkbeiner,
  arXiv:1206.1616 [astro-ph.HE].









\bibitem{newmodels}
  A.~Ibarra, S.~Lopez Gehler and M.~Pato,
  arXiv:1205.0007 [hep-ph];
  E.~Dudas, Y.~Mambrini, S.~Pokorski and A.~Romagnoni,
  arXiv:1205.1520 [hep-ph];
  J.~M.~Cline,
  arXiv:1205.2688 [hep-ph];
  K.~-Y.~Choi and O.~Seto,
  arXiv:1205.3276 [hep-ph];
  B.~Kyae and J.~-C.~Park,
  arXiv:1205.4151 [hep-ph];
  B.~S.~Acharya, G.~Kane, P.~Kumar, R.~Lu and B.~Zheng,
  arXiv:1205.5789 [hep-ph];
  M.~R.~Buckley and D.~Hooper,
  Phys.\ Rev.\ D {\bf 86} (2012) 043524
  [arXiv:1205.6811 [hep-ph]];
  D.~Das, U.~Ellwanger and P.~Mitropoulos,
  JCAP {\bf 1208} (2012) 003
  [arXiv:1206.2639 [hep-ph]];
  Z.~Kang, T.~Li, J.~Li and Y.~Liu,
  arXiv:1206.2863 [hep-ph];
  J.~-C.~Park and S.~C.~Park,
  arXiv:1207.4981 [hep-ph];
  S.~Tulin, H.~-B.~Yu and K.~M.~Zurek,
  arXiv:1208.0009 [hep-ph];
  T.~Li, J.~A.~Maxin, D.~V.~Nanopoulos and J.~W.~Walker,
  arXiv:1208.1999 [hep-ph];
  J.~M.~Cline, A.~R.~Frey and G.~D.~Moore,
  arXiv:1208.2685 [hep-ph];
  Y.~Bai and J.~Shelton,
  arXiv:1208.4100 [hep-ph];
  L.~Bergstrom,
  arXiv:1208.6082 [hep-ph];
  L.~Wang and X.~-F.~Han,
  arXiv:1209.0376 [hep-ph].

\bibitem{lp2}
  H.~M.~Lee, M.~Park and W.~-I.~Park,
  arXiv:1205.4675 [hep-ph].


\bibitem{tait}
  A.~Rajaraman, T.~M.~P.~Tait and D.~Whiteson,
  arXiv:1205.4723 [hep-ph];



\bibitem{background}
  S.~Profumo and T.~Linden,
  JCAP {\bf 1207} (2012) 011
  [arXiv:1204.6047 [astro-ph.HE]];
  A.~Boyarsky, D.~Malyshev and O.~Ruchayskiy,
  arXiv:1205.4700 [astro-ph.HE].





\bibitem{pamela}
  O.~Adriani {\it et al.}  [PAMELA Collaboration],
  Phys.\ Rev.\ Lett.\  {\bf 105} (2010) 121101
  [arXiv:1007.0821 [astro-ph.HE]].

\bibitem{continuum}
  X.~Chu, T.~Hambye, T.~Scarna and M.~H.~G.~Tytgat,
  arXiv:1206.2279 [hep-ph];
  W.~Buchmuller and M.~Garny,
  JCAP {\bf 1208} (2012) 035
  [arXiv:1206.7056 [hep-ph]];
  T.~Cohen, M.~Lisanti, T.~R.~Slatyer and J.~G.~Wacker,
  arXiv:1207.0800 [hep-ph];
  I.~Cholis, M.~Tavakoli and P.~Ullio,
  arXiv:1207.1468 [hep-ph];
  X.~-Y.~Huang, Q.~Yuan, P.~-F.~Yin, X.~-J.~Bi and X.~-L.~Chen,
  arXiv:1208.0267 [astro-ph.HE];
  R.~Laha, K.~C.~Y.~Ng, B.~Dasgupta and S.~Horiuchi,
  arXiv:1208.5488 [astro-ph.CO].
  
 

 


\bibitem{wagner}
  A.~Joglekar, P.~Schwaller and C.~E.~M.~Wagner,
  arXiv:1207.4235 [hep-ph];
  M.~Carena, I.~Low and C.~E.~M.~Wagner,
  arXiv:1206.1082 [hep-ph].


\bibitem{AH}
  N.~Arkani-Hamed, K.~Blum, R.~T.~D'Agnolo and J.~Fan,
  arXiv:1207.4482 [hep-ph].



\bibitem{almeida}
  L.~G.~Almeida, E.~Bertuzzo, P.~A.~N.~Machado and R.~Z.~Funchal,
  arXiv:1207.5254 [hep-ph].


\bibitem{weiner}
  J.~Kearney, A.~Pierce and N.~Weiner,
  arXiv:1207.7062 [hep-ph].







\bibitem{su3w}
  S.~Weinberg,
  Phys.\ Rev.\ D {\bf 5} (1972) 1962.


\bibitem{su3w2002}
  S.~Dimopoulos and D.~E.~Kaplan,
  Phys.\ Lett.\ B {\bf 531} (2002) 127
  [hep-ph/0201148];
  T.~-j.~Li and W.~Liao,
  Phys.\ Lett.\ B {\bf 545} (2002) 147
  [hep-ph/0202090];
  S.~Dimopoulos, D.~E.~Kaplan and N.~Weiner,
  Phys.\ Lett.\ B {\bf 534} (2002) 124
  [hep-ph/0202136];
  H.~-D.~Kim, J.~E.~Kim and H.~M.~Lee,
  JHEP {\bf 0206} (2002) 048
  [hep-th/0204132].


\bibitem{ko}
  S.~-W.~Baek, P.~Ko and W.~-I.~Park,
  JHEP {\bf 1202} (2012) 047
  [arXiv:1112.1847 [hep-ph]].


\bibitem{wbosonmass}
  T.~Aaltonen {\it et al.}  [CDF Collaboration],
  Phys.\ Rev.\ Lett.\  {\bf 108} (2012) 151803
  [arXiv:1203.0275 [hep-ex]];
  V.~M.~Abazov {\it et al.}  [D0 Collaboration],
  Phys.\ Rev.\ Lett.\  {\bf 108} (2012) 151804
  [arXiv:1203.0293 [hep-ex]].

\bibitem{mtrott}
  J.~R.~Espinosa, C.~Grojean, M.~Muhlleitner and M.~Trott,
  arXiv:1207.1717 [hep-ph].


\bibitem{deltaT}
  L.~Lavoura and J.~P.~Silva,
  Phys.\ Rev.\ D {\bf 47} (1993) 2046.




\bibitem{djouadi}
  J.~Baglio, A.~Djouadi and R.~M.~Godbole,
  arXiv:1207.1451 [hep-ph].




















\bibitem{ramond}
  H.~Arason, D.~J.~Castano, B.~Keszthelyi, S.~Mikaelian, E.~J.~Piard, P.~Ramond and B.~D.~Wright,
  Phys.\ Rev.\ D {\bf 46} (1992) 3945.





\bibitem{BBN}
  M.~Pospelov,
  Phys.\ Rev.\ Lett.\  {\bf 98} (2007) 231301
  [hep-ph/0605215];
  J.~Pradler and F.~D.~Steffen,
  Phys.\ Lett.\ B {\bf 666} (2008) 181
  [arXiv:0710.2213 [hep-ph]];
  S.~Bailly, K.~Jedamzik and G.~Moultaka,
  Phys.\ Rev.\ D {\bf 80} (2009) 063509
  [arXiv:0812.0788 [hep-ph]].


\bibitem{colliders}
  S.~Chatrchyan {\it et al.}  [CMS Collaboration],
  Phys.\ Lett.\ B {\bf 713} (2012) 408
  [arXiv:1205.0272 [hep-ex]];
ATLAS Note, ATLAS-CONF-2012-075.



\end{thebibliography}
\end{document}